\documentclass[usenatbib]{mnras}
\usepackage{siunitx}
\usepackage{graphicx}	
\usepackage[utf8]{inputenc}
\usepackage[english]{babel}
\usepackage{float}
\usepackage{amssymb, amsmath, yhmath}
\usepackage{verbatim}

\renewcommand{\d}[1]{\! \mathrm{d}#1 \:}
\newcommand{\deriv}[2]{\frac{\d{#1}}{\d{#2}}}
\newcommand{\pderiv}[2]{\frac{\partial{#1}}{\partial{#2}}}

\newcommand{\sech}{\mathrm{sech}}
\newcommand{\EV}[1]{\left< #1 \right>}

\setcitestyle{authoryear, open={(},close={)}}

\usepackage{graphicx}

\newcommand{\squote}[1]{\lq #1\rq}

\renewcommand{\d}[1]{\ensuremath{\operatorname{d}\!{#1}}}
\newcommand{\poweV}[1]{\SI{e#1}{\electronvolt}}

\newcommand{\lcdm}{$\Lambda$CDM}
\DeclareSIUnit\parsec{pc}
\DeclareSIUnit\lightyear{ly}
\DeclareSIUnit\year{yr}
\begin{document}

\title[Dark Matter Heating]{Heating of Milky Way disc Stars by Dark Matter Fluctuations in Cold Dark Matter and Fuzzy Dark Matter Paradigms}
\author[B. V. Church et al.]{
Benjamin V. Church$^{1}$ \thanks{Contact e-mail: bvc2105@columbia.edu}, Philip Mocz$^{2}$ \thanks{Einstein Fellow},
Jeremiah P. Ostriker$^{1 \, 2}$ \thanks{Columbia Contact e-mail: jpo@astro.columbia.edu 
\newline
Princeton Contact e-mail: ostriker@princeton.edu} 
\\
$^{1}$Columbia University, Department of Astronomy, New York, NY 10025, USA.
\\
$^{2}$Department of Astrophysical Sciences, Princeton University, 4 Ivy Lane, Princeton, NJ, 08544, USA.}
\date{Accepted to MRNAS, Feburary 15, 2019}
\maketitle
\begin{abstract}
Although highly successful on cosmological scales, Cold Dark Matter (CDM) models predict unobserved over-dense \squote{cusps} in dwarf galaxies and overestimate their formation rate. We consider an ultra-light axion-like scalar boson which promises to reduce these observational discrepancies at galactic scales. The model, known as Fuzzy Dark Matter (FDM), avoids cusps, suppresses small-scale power, and delays galaxy formation via macroscopic quantum pressure. We compare the substructure and density fluctuations of galactic dark matter haloes comprised of ultra-light axions to conventional CDM results. Besides self-gravitating subhaloes, FDM includes non-virialized over-dense wavelets formed by quantum interference patterns which are an efficient source of heating to galactic discs. We find that, in the solar neighborhood, wavelet heating is sufficient to give the oldest disc stars a velocity dispersion of $\sim \SI{30}{\kilo\meter\per\second}$ within a Hubble time if energy is not lost from the disc, the velocity dispersion increasing with stellar age as $\sigma_D \propto t^{0.4}$ in agreement with observations. 
Furthermore, we calculate the radius-dependent velocity dispersion and corresponding scale height caused by the heating of this dynamical substructure in both CDM and FDM with the determination that these effects will produce a flaring that terminates the Milky Way disc at $\SIrange{15}{20}{\kilo \parsec}$. Although the source of thickened discs is not known, the heating due to perturbations caused by dark substructure cannot exceed the total disc velocity dispersion. Therefore, this work provides a lower bound on the FDM particle mass of $m_a > \SI{0.6 e-22}{\electronvolt}$. Furthermore, FDM wavelets with this particle mass should be considered a viable mechanism for producing the observed disc thickening with time.
\end{abstract}


\section{Introduction}

\subsection{Background on Dark Matter}

The dark sector contributes a sizable majority of all matter in the universe. The existence of one or more kinds of dark matter is essential to galaxy formation and consistency with observed disc rotation curves. Although ``dark'' to direct electromagnetic observation, irregularities in the structure of that dark matter can have observable dynamical effects on the visible stars. Over-dense regions of dark matter collapse to roughly localized concentrations of matter called ‘haloes’ which are well-described by approximately spherically symmetric density profiles \citep{structure}. However, similar haloes may vary significantly from these profiles in small-scale power i.e. substructure. Various models for the underlying physics of dark matter predict various patterns of density irregularities which, though invisible to the electromagnetic spectrum, are capable of perturbing the orbits of stars via gravitational effects. We will attempt to estimate those effects on the observable stars in galactic discs to better constrain our understanding of the fundamental nature of the dark matter.
             
\par

The standard cosmological model (\lcdm) contains a massive electromagnetically non-interacting component known as Cold Dark Matter (CDM) with density parameter $\Omega_c = 0.259 \pm 0.006$ \citep{planck}. The properties of this substance are largely unknown and are assumed to be approximately that of a perfect fluid with negligible pressure compared to its energy density i.e. comprised of cold or non-relativistic matter at high redshift. CDM simulations suggest that substructure forms self-gravitating clumps which are qualitatively similar to free dark matter haloes. We adopt a substructure formalism which tracks subhaloes as an estimate of total substructure. The secondary gravitational effects of these subhaloes is dependent on the subhalo mass function and the shape of small halo profiles, which is directly dependent on properties of the dark matter physics under consideration. Constraints obtained from The Sloan Digital Sky Survey measurement of Ly-$\alpha$ forest power spectra and surveys of dwarf galaxies below the free-streaming scale rule out most models of relativistic (hot) matter as the primary component of dark matter \citep{can_neutrinos}. Attempts to directly detect CDM particles in various plausible models of physics beyond the standard model have so far been without success \citep{direct_detection}.

\subsection{Apparent Problems for the CDM Scenario}

Although amazingly accurate on cosmological scales, CDM models fail to make accurate predictions when compared with observations at distance scales less than \SI{1}{\kilo\parsec}. When compared with studies of dwarf galaxy rotation curves, CDM predicts unobserved density \squote{cusps} in the  centres of dark matter haloes \citep{ultralight}. This failure is known as the \squote{cusp-core problem}. Another serious concern is the \squote{missing satellite problem}  \citep{missing_satellites}. The number of satellite galaxies predicted for a Milky-Way-like galaxy is greater than what we observe by an order of magnitude. This issue is sharpened by the \squote{too big to fail} problem of galaxy formation that claims some of the predicted satellites are so massive that it is difficult to imagine that they would not form any stars \citep{too_big_to_fail}. Various complex phenomena have been suggested as solutions to these problems such as baryonic feedback mechanisms which redistribute matter or  hypothetical dark matter self-interactions.

\subsection{Ultra-light Dark Matter}

An alternative model being studied known as Fuzzy Dark Matter (FDM) characterizes the bulk of dark matter as comprised of ultra-light (with masses in the range $\poweV{-24} \leq m_a \leq \poweV{-18}$) bosons whose characteristic de Broglie wavelength is on the order of $\SI{1}{\kilo\parsec}$. Recent reviews of the astrophysical properties of the FDM alternative can be found in \cite{ultralight} and \cite{axion_cosmology}. Such ultra-light particles are possible in various theories beyond the standard model \citep{axion_cosmology}. In particular, the class of axion-like particles is a perfect candidate for FDM and therefore, we refer to the mass of the constituent particles of FDM as the axion mass $m_a$. Macroscopic quantum mechanical wave effects “smear” the density profile on scales less than $\sim \SI{1}{\kilo\parsec}$ which removes problematic cusps from the centres of haloes. These cusps are replaced by dense self-gravitating quantum states known as soliton cores which may facilitate indirect observational evidence for FDM. Furthermore, the quantum mechanical pressure caused by the large wavelength of FDM suppresses the formation of low-mass haloes and entirely eliminates the formation of any halo or subhalo smaller than this wavelength, reducing or eliminating the missing satellites problem \citep{substructure_FDM}. Significant further effort remains in determining further discrepancies between FDM and CDM and whether these can, in conjunction with observational constraints, rule out one or both of these models.
 
\subsection{Dynamical Effects of Dark Matter on Stars}

The dynamical effects of CDM have been studied in some detail, especially regarding the accretion and tidal stripping of CDM subhaloes by large galaxies and their associated dark matter haloes. On the other hand, research on the dynamics of FDM distributions is developing rapidly (c.f. \cite{schive_solitons, Schrodinger-Poisson, Schive-virialized-wave-halos, relaxation}) but still is in its infancy. However, dynamical studies have focused on simulations which inherently have a fixed resolution and thus are limited to small (approximately \SI{1}{\mega \parsec} scales) cosmological volumes and low ($M \ll 10^{12} M_{\odot}$) halo masses. In this paper, we present analytic estimates for dynamical density fluctuations which do not suffer from limited resolution. The resolution limit of simulations causes underestimates of subhalo populations and therefore to their dynamical effects. 

\par 
	CDM and FDM models predict significant differences in the distribution of subhalo masses. While CDM predicts a power law divergence in the number of smaller and smaller subhaloes \citep{power_spectra_JPO}, the FDM subhalo mass function approaches zero below a cutoff scale determined by the axion mass expressed by the Jean’s scale at which quantum pressure and gravitational attraction balance:
\setlength{\belowdisplayskip}{4pt} \setlength{\belowdisplayshortskip}{4pt}
\setlength{\abovedisplayskip}{4pt} \setlength{\abovedisplayshortskip}{4pt}

\begin{equation}
k_J = 66.5 \cdot (1+z)^{1/4} \left( \frac{\Omega_a h^2}{0.12} \right) \left(\frac{m_a}{\poweV{-22}} \right)^{1/2} \si{\per\mega\parsec},
\end{equation}
\noindent
as given by \citet{axion_cosmology}. Subhaloes and primary haloes less massive than $\sim \SI{e9}{M_{\odot}}$ (at $z = 0$) are thus suppressed for axion masses $m_a < \SI{e-22}{\electronvolt}$. For ultra-light axions, the subhalo mass function is significantly suppressed through the entire range of substructure for a halo comparable to the Milky Way’s. However, FDM subhaloes will exhibit solitons which, for moderately-sized subhaloes and large axion masses, are very dense and largely unaffected by tidal disruption. 

\par 

The main deviation from CDM phenomena occurs on scales smaller than the coherence length. Above this scale, the Schr\"{o}dinger-Poisson–Vlasov-Poisson correspondence predicts that the self-gravitating non-linear Schr\"{o}dinger equation governing FDM reproduces the physics of collisionless particulate dark matter \citep{Schrodinger-Poisson}. Simulations using zoom-in techniques are able to resolve complex structure in the core of an FDM halo and have discovered large-amplitude oscillations in the core density of such haloes \citep{structure-FDM-halos}. In addition, a surprising standing wave phenomenon has been observed in the density profiles produced by small-scale FDM computer simulations \citep{cold_and_fuzzy}. These standing waves, named wavelets, radically alter the dark matter profile and may, through gravitational interactions, disturb the baryonic components in directly measurable ways. The primary effect of these wavelets is to introduce time-varying perturbations to the gravitational potential on a much shorter time-scale than the evolution of primary structure after the system has come into virial equilibrium \citep{Schrodinger-Poisson}. 
\par
	The velocity dispersion of stars in a galactic disc provides a proxy to study the gravitational perturbations produced by the dark matter density fluctuations. Observations of spiral galaxies at increasing redshift show that velocity dispersion increases with age via an approximate power law $\sigma \propto t^\beta$ where $\beta \approx \frac{1}{3}$ \citep{heating_history}. Therefore, the best observational bounds on the amount of gravitational  fluctuations due to dynamical perturbations will come from the analysis of old, thick disc stars. We will see that primary substructure in both CDM and FDM -- self-gravitating subhaloes -- will heat and thicken the Milky Way disc exterior to the solar orbit causing extreme flaring by $\SIrange{15}{20}{\kilo\parsec}$. In addition, the gravitational fluctuations produced by FDM wavelets provide significant heating at and interior to the $\sim \SI{8}{\kilo\parsec}$ radius of the sun. If the particle mass if on the order of $\SI{e-22}{\electronvolt}$, this process can explain the nature of the old thick disc. However, a still lower particle mass can be excluded as it would have heated and thickened the local disc by more than is observed.  	

\section{Galactic discs}

\hspace{5mm} Current observations suggest that the discs of spiral galaxies may be roughly modeled by an approximately exponential profile in radius and in vertical structure. The vertical scaling is determined by the mass density of the disc and the perpendicular velocity dispersion.  The bulk of the stars in normal discs are divided into a cold `thin disc' of relatively young stars and a hot `thick disc' of old stars \citep{binney_tremaine_2008}. 

\par

The identification of and first evidence for the presence of a thick disc in the Milky Way Galaxy was presented by \cite{milkyway-thickdisc} who showed that the  density profile of stars as a function of distance from the galactic plane could not be explained by a single exponential, but rather by two exponentials with scale heights of $\SI{300}{\parsec}$ and $\SI{1350}{\parsec}$ respectively.
The vertical velocity dispersion and therefore scale height of the old stars is significantly greater, which is why this population is referred to as the `thick disc'. Stars in the thick disc are very old, most with ages exceeding \SI{12}{\giga \year} and the majority of thick disc sub-giants dating from 12 - \SI{13}{\giga \year} ago \citep{age_of_thick_disc}. In comparison, the stars comprising the thin disc have a much wider range of ages, most younger than $\SI{8}{\giga \year}$. 
Subsequent studies have established chemical differences in metallicity and $\alpha$-abundance between the stars in the thin and thick discs with the thin disc consisting of the younger high metallicity population I stars \citep{GALAH_thick_disc, chemical_thick_disc}. Although the thick disc is the older of the two, it ought not be confused with the \textit{old} disc which refers to the older component stars of the \textit{thin} disc. Such terminology is used because the thin disc is comprised of stars with a broad range of ages.  
\par
Studies of the velocity dispersion of Milky Way stars at varying angles and scale heights showed that the velocity dispersion of the thick disc is age dependent and fit approximately by a power law $\sigma_D \propto t^{\beta}$ where the best estimates for the exponent are $\beta \approx \frac{1}{3}$ \citep{heating_history}. The origin of this disc structure is unknown but there are three widely considered explanations. One hypothesis is that the thick disc formed first with subsequent stellar generations born in ever slimmer discs due to energy loss of the disc to gas shocks. However, this explanation struggles to explain the profile of observed decrease in velocity dispersion with increasing redshift \citep{emergence-thick-disc}. An alternative hypothesis is that tidal interactions between galaxy close encounters can inject energy into existing stars which effectively puffs up the disc \citep{thick-disc-mergers}. \cite{mergers} ruled our major mergers in the accretion history of the Milky Way which are inconsistent with the existence of a cold disc and further limit the total mass of stars or dark matter which could have been accreted to at most 4\% of the mass of the galaxy (interior to $R_{\odot}$) over the lifetime of the Milky Way. Furthermore, estimates of the actual rate of accretion on to spiral galaxies are consistent with this limit and large enough for the accretion of dwarf satellites to be a viable mechanism explaining observed disc thickness \citep{mergers}.
The recent study by \cite{gaia_normal_process} based on Gaia data revives the proposal that the thick disc was produced by the ancient accretion of a Small Magellanic Cloud-like satellite system. This otherwise attractive possibility would not produce the age-velocity dispersion observed in the Milky Way thick disc. 
A final possibility, and the mechanism we consider in this paper, is that the thick disc is formed by continually heating older stars \citep{thin-and-thick-disc}. 
\par 
Assuming that galactic disc components are formed by continual heating, the primary source of this heating remains uncertain. Interactions between spiral modes and radial migration of stars due to spiral arms have been proposed as a mechanism for heating spiral discs \citep{radial_migration}. However, the ability to heat up to the high velocity dispersion of the old disc stars by spiral modes remains to be demonstrated. The time-varying quadrupole moment of the bar of a barred spiral galaxy can also drive vertical disc heating. \cite{vertical_heating_modes} show that, in cosmological-zoom calculations of Milky Way-like CDM haloes, rotating bars are the dominant effect while spiral density waves and radial migration are sub-dominant. Furthermore, they find that dynamical subhaloes and satellites with masses above $10^{10} M_{\odot}$ dominate all the above disc-disc interactions. 
Although spiral arms couple weakly to the vertical motions of stars, such effects are able to efficiently increase their planar velocity distribution which, in turn, are converted into vertical motions by giant molecular clouds (GMCs) acting as scattering centres \citep{vertical_structure_and_GMC}. Combined GMC and spiral/bar heating is successful in explaining the age-velocity dispersion relation for the galactic thin disc but is unable to produce the thick disc of a Milky Way-like galaxy \citep{heating_history}.
Here we will consider the possibility that disc heating is caused by interactions with dynamical dark halo substructure including both subhaloes and FDM wavelets. In this case, we will be able to sharply restrict the parameter space of possible dark matter paradigms. Regardless of the origin of thick discs, the heating due to dark substructure can at most account for the total observed velocity dispersion of the thick disc. Such a relaxation rate is proportional to the mass of the perturbers producing impulsive heating (which, in the case of FDM, should not be confused with the mass of the ultra-light particles comprising them). As we shall show, FDM wavelets dominate other sources of heating from dark matter fluctuations at small radii. Furthermore, the characteristic wavelength of dark interference patterns determining the mass of the wavelet perturbers is inversely related to the FDM particle mass making the mass of these perturbers scale as the inverse cube of the particle mass. Thus to avoid overheating the disc, these results put a lower bound on the mass of the FDM axion which is agnostic to the mechanism behind producing thick discs. 

\par

The discs of normal spiral galaxies approximately follow an exponential radial density profile,
\begin{equation}
\rho(r, z) = \frac{\Sigma_0}{2 h(r)} e^{-r/r_0} \mathrm{sech}^2{(z/h(r))}, 
\end{equation}
where $r_0$ is the scaling radius and $h(r)$ is the scale height at a given radius. The indicated vertical dependence is exact only for a disc which is isothermal along vertical slices of constant radius. Furthermore, the projection of the mass density on to the galactic plane has the same exponential form. In particular, the local surface density as a function of radius has approximately the form,
\begin{equation}
\Sigma(r) = \int_{-\infty}^{\infty} \rho(r, z) \d{z} = \Sigma_0 e^{-r / r_0}.
\end{equation}
For the Milky Way the local surface density is estimated to be $\Sigma_{\odot} = \SI{68}{M_{\odot} \per \parsec \squared}$ using dynamical measurements \citep{dynamical_measurement}. We take the Milky Way disc to have an exponential surface density profile with $r_0 \approx \SI{3.2}{\kilo\parsec}$ to fit the estimated component-wise rotation curve of \cite{milky_way_halo}. The vertical profile is determined by hydro-static equilibrium between the gravity from the planar density distribution of the disc and supporting pressure given by the vertical velocity dispersion $p = \rho \sigma_D^2$ where $\sigma_D$ is the velocity dispersion of the disc. The density profile must therefore satisfy,
\begin{equation}
- 4 \pi G \rho = \frac{\partial^2 \phi}{\partial z^2} = \pderiv{}{z} \left( \frac{1}{\rho} \pderiv{ \rho \sigma_D^2}{z} \right).
\end{equation}  
Integrating through the disc and approximating $\sigma_D^2$ as varying slowly vertically through the disc we arrive at the following expression relating the scale height and vertical velocity dispersion of the disc:
\begin{equation} \label{scale}
h(r) = \frac{\sigma_D(r)^2}{\pi G \Sigma(r)}.
\end{equation}
Stars oscillating vertically through the disk at a fixed radius are acted on by a gravitational acceleration and corresponding gravitational potential,
\begin{subequations}
\begin{align}
-\mathbf{g}(z) & = 4 \pi G \int_0^z \rho(r, z') \d{z'} = 2 \pi G \Sigma \tanh{(z/h)}.
\\
\phi(z) & = - \int_0^z \mathbf{g}(z') \d{z'} = 2 \pi G \Sigma \: h \: \log{\cosh{(z / h)}}.
\end{align}
\end{subequations}
Thus, the vertical oscillations of such a star with maximum vertical height $z_m$ have a period given by,
\begin{subequations}
\begin{align}
\frac{P}{2} & = \int_{-z_m}^{z_m} \frac{\d{z'}}{\sqrt{2(\phi(z_m) - \phi(z'))}} = \sqrt{\frac{h}{4 \pi G \Sigma}} I_P(z / h) ;
\\
I_P(y) &= \int_{-y}^y \frac{\d{x}}{\sqrt{\log{\cosh{y}} - \log{\cosh{x}}}}.
\end{align}
\end{subequations}
The function $I_P$ is monotonically increasing and bounded below by $\pi \sqrt{2}$. 
Using the disc scale height to velocity dispersion relation given by equation \eqref{scale}, the vertical oscillation period can be written as,
\begin{equation} \label{period}
P = \frac{\sigma_D}{\pi G \Sigma} I_P(z_m / h) = \frac{h}{\sigma_D} I_P(z_m / h)
\end{equation}
The period of disc oscillations determines the adiabatic cutoff for heating due to perturbing interactions. Perturbations slower than the vertical oscillation period only produce adiabatic changes to the orbits of stars which impart negligible heating to the disc. We assume that disk stars are born with the velocity dispersion of the gas and their velocity profile evolves from there due to heating. The extremities of the thick disk is formed from stars which evolved from the small proportion of stars with large $z_m$ on the order of $2$ to $3$ scale heights. Under the assumption that disc thickening is due primarily to the proposed dark substructure dynamics, the dependence on radius of the disc velocity dispersion $\sigma_D$ can be calculated explicitly (see section \ref{secton:heating}). Using the above relations and the radial dependence of $\sigma_D$ we can predict the radial thickness profile of the galactic disc and compare these results to measurements of the flaring of actual discs.  
\par
We now turn our attention to the Milky Way which has the best studied galactic disc. \cite{milky_way} has performed a comprehensive study of the various distribution functions of Milky Way stars and, in particular, has compiled data on the velocity dispersion of the thin and thick discs as a function of vertical distance. \cite{milky_way} proposes a ``pseudo-isothermal'' distribution function for the velocity dispersion which has the form,
\begin{equation}
f_z(J_z) = \frac{\left( \Omega_z J_z + V_\gamma^2 \right)^{-\gamma}}{2 \pi \int_{0}^{\infty} \d{J_z} \: \left( \Omega_z J_z + V_\gamma^2 \right)^{-\gamma} },
\end{equation}  
in term of the action variable $J_z$. For stars in the solar neighborhood with $r \approx R_{\odot} =  \SI{8.3}{\kilo \parsec}$ the best-fitting parameters of $\gamma = 2.6$ and $V_\gamma = \SI{18.7}{\kilo \meter \per \second}$ give excellent agreement with survey data collected by \cite{milkyway-thickdisc}. Using this formalism, the main component of the perpendicular velocity dispersion of the thick disc is $\sigma_D = \SI{32 \pm 1}{\kilo \meter \per \second}$ at the neighborhood near the solar radius $R_{\odot}$.

\section{The Theory of Heating via Substructure Dynamics} \label{secton:heating}

Here, we present a model for how dynamical substructure of a dark matter halo can heat galactic discs due to time-varying gravitational interactions. We furthermore discuss the evolution of the velocity dispersion and disc profile in the presence of such heating. 

\subsection{Estimating The Heating Due to Transits}

We first provide a rough analysis of the heating due to transiting of over-dense regions of dark matter. Each transit causes, on average, a change in velocity of approximately,
\begin{equation}
\EV{\Delta v^2} = \left( \frac{M_l G}{b v_l} \right)^2,
\end{equation}
where $M_l$ is the mass of the perturbing object, $v_l$ is its velocity, and $b$ is the distance at closest approach and we have assumed that the perturbing object is much more massive than the test particle whose velocity changes have been estimated. As usual, we neglect terms of the form $\EV{\bf{v} \cdot \Delta \bf{v}}$ which average to zero under moderate assumptions about spatially uncorrelated velocities and perturbing impulses. However, the heating is ineffective if these perturbations are adiabatic (i.e. slow compared to the oscillation period of the star in question) which will merely cause gradual periodic changes rather than overall heating to the orbits whose adiabatic invariants remain fixed. However, if the heating is roughly impulsive then the orbits will be irreversibly perturbed. The non-adiabatic condition approximately corresponds to the constraint,
\begin{equation} \label{adiabatic}
\frac{b}{v_l} < \frac{P}{2},
\end{equation}
where $P$ is the characteristic vertical period of the objects subjected to perturbation, in this case, the period for vertical oscillations through the galactic disc. The heating of the disc, denoted by $\mathcal{H}$, is measured by tracking impulsive changes to the one-dimensional velocity dispersion of the disc $\sigma_D^2 = \left< \dot{z}^2 \right>$ over time. That is, up to factors of one-half the mass of the stars, the heating measures the rate that fluctuations pump energy into the disc. Note that $\mathcal{H} \neq \deriv{\sigma_D^2}{t}$ (perhaps with the derivatives averaged over a suitable interaction time-scale to smooth impulsive interactions) because the former quantity only takes into account energy delivered to the disc and does not include adiabatic changes to the disc structure which induce an overall time dependence in the velocity dispersion. For example, as the disc is heated due to the aforementioned transiting fluctuations, the disc will ``puff up'' in scale height transferring the delivered kinetic energy to additional gravitational potential energy. Keeping in mind this distinction, the total rate of heating is therefore given by the accumulation of all such encounters,
\begin{subequations}
\begin{align}
\mathcal{H} = \deriv{\sigma_D^2}{t} \bigg|_{\text{impulsive}} &= \int_{b_{\min}}^{b_{\min}} (2 \pi b) \: \d{b} \: n v_l \: \left( \frac{M_l G}{b v_l} \right)^2 ;
\\
& = \frac{M_l^2 G^2}{v_l} \: 2 \pi n \log{\Lambda},
\end{align}
\end{subequations}
where we have defined,
\begin{equation}
\log{\Lambda} \equiv \log{\left( \frac{b_{\text{max}}}{b_{\text{min}}} \right)},
\end{equation}
which is the gravitational equivalent of the Coulomb logarithm. Here $v_l$ is the relative velocity between the disc and the perturbing objects in the halo which have velocity dispersion $\sigma_H$. The maximum distance across which the heating is efficient, $b_{\max}$, is fixed by equations \eqref{adiabatic} and \eqref{period} and the minimum distance, $b_{\min}$, is given by the characteristic size for the perturbing objects denoted $r_{l}$. 
Therefore, in terms of the overall one-dimensional velocity dispersion, we derive the time-dependent heating as,
\begin{subequations}
\begin{align}
\mathcal{H} &= \frac{M_l^2 G^2}{\sigma_H} \: 2 \pi n \log{\Lambda} ;
\\
\Lambda & = \frac{\sigma_H \sigma_D}{2 \pi G \Sigma \: r_l} I_P(z_m / h).
\end{align}
\end{subequations}
A more detailed calculation found in \cite{milkywayblackholes} gives somewhat altered numerical factors,
\begin{align} \label{heating}
\mathcal{H} = \frac{8 \pi M_l^2 G^2}{\sqrt{2} \sigma_H} \: n \: \log{\Lambda}.
\end{align}
An alternative approach for calculating heating due to transits is given by \cite{ultralight} which only considers the tidal effects on disc heating. That model assumes that only local dispersion acting across a characteristic scale much smaller than $h$ can lead to overall heating. However if processes such as outgoing density waves are an inefficient means for dissipating differential motions of the disc then dynamical friction between oscillating disc components will tend to dissipate oscillation energy into disc heating and therefore contribute to the thickening. In this analysis, we will assume that such processes are inefficient and therefore equation \eqref{heating} gives a reasonable estimate of the total rate of disc heating. An equivalent formulation is that our method assumes that all energy delivered to disc stars by dynamical perturbations contributes, eventually, to heating the disc and that short wavelength bending waves induced by anisotropic interactions with dark matter perturbers are ultimately damped out such that their energy is deposited in stellar random motions.

\par

Thus, the calculation we provide is based on the total kinetic energy delivered into vertical motions of the disc. These motions can be broken in two components: those reflecting a direct increase in the z-velocity dispersion and those invested in bending waves which give one part of the disc a z-velocity with respect to the mean disc velocity. Some fraction of the second component could be damped by transfer of energy back to the dark matter, to gas motions, or be transferred radially outwards in propagating bending disc waves. Thus, our calculation provides an upper bound on the disc heating by subhaloes and wavelets with the calculation given by equation (37) of \cite{ultralight} giving a lower bound. 
     
\subsection{Heating Due to Subhaloes}
\subsubsection{Spatial Profile}

We model self-gravitating substructure as comprised entirely of subhaloes which act as distinct massive particles subject only to gravitational attraction and tidal disruption. Following \citet{tidal_limit} and \cite{unified_model} we adopt a
simplified model of subhalo
formation and dynamics. The initial shape of the subhalo mass function is assumed to
be spatially invariant, i.e. the position
and (density normalized) mass variables are decoupled. We would expect the unresolved mass function to have a fixed shape since it is determined entirely from the dark matter power spectrum \citep{unified_model} via some variant of the Press-Schechter formalism \citep{press-schechter}. 
In particular, the CDM power spectrum exhibits self-similarity describing the incorporation of smaller subhaloes into larger haloes. 
The total density of subhaloes is thus assumed to initially
follow the Navarro--Frenk--White (NFW) profile:
\begin{equation}
\rho(r) = \frac{\rho_0}{r/r_c (1+r/r_c)^2},
\end{equation} 
which describes the background 
density profile of the
primary dark matter halo given by \citet{structure} from CDM simulations. This profile is fairly universal for the haloes of massive cold particles which we assume accurately describe subhaloes before they are subject to tidal processes. Furthermore, we are working under the assumption that the primary modifications of subhaloes relative to free haloes arise from tidal stripping. Wave dark matter models modify the primary halo density profile away from NFW. The core behavior in FDM is primarily due to quantum pressure which does not act between self-gravitating lumps. Therefore, we do not expect the spatial distribution of subhalo formation to be significantly altered from FDM. That said, due to extreme tidal disruption, subhaloes near the core of the primary halo contribute negligibly, so the exact shape of their distribution near the core is unimportant.  

\subsubsection{The Subhalo Mass Function}

The only additional information needed to calculate the perturbations due to subhaloes is the distribution of subhalo populations as a function of mass. However, the tidal interactions between the primary halo and its interior subhaloes complicates the elegant halo mass functions determined by the over-density correlation function and power spectrum. We approximate the subhalo populations by assuming that the unresolved subhalo mass function has the same form as the free halo mass function derived from the extended Press--Schechter formalism \citep{press-schechter, excursion-set}. The unresolved mass function is then modified by tidally truncating unresolved haloes such that the unresolved halo mass function is shifted towards lower mass and the modified subhalo mass function becomes coupled in subhalo mass and radius inside the primary halo. Therefore, the population of subhaloes is entirely determined by two functions, $\frac{\d{n}}{\d{\log{M}}}$, the unresolved (free) halo mass function, and $T_R(M)$, the truncated mass function at a radius $R$. 

\subsubsection{Calculating the Heating}

In terms of these functions, the rate of heating is given by,
\begin{equation} \label{subhaloheating}
\mathcal{H} = \frac{8 \pi G^2}{\sqrt{2} \: \sigma_H} \int_0^{M} T_R(m)^2  \deriv{n}{m}  \log{\left( \frac{\sigma_H \sigma_D}{2 \pi G \Sigma \: r_{\text{t}}(m)} \right)} \d{m},
\end{equation}   
where $r_{\text{t}}(m)$ is the radius of a subhalo of mass $m$ after truncation at a radius $R$. 
If we assume that the primary halo and substructure formed quickly on the scale of the lifetime of the disc, and that dark substructure dynamics in the form of transiting subhaloes alone constitutes the primary source of disc thickening, then equation \eqref{subhaloheating} gives the time-dependence of $\sigma_D$ over the entire history of the halo. Furthermore, the rate of heating given by equation \eqref{subhaloheating} depends on $\sigma_D$ only logarithmically. Therefore, the square of the velocity dispersion increases at an approximately constant rate analogous to a random walk. Naively, this heating model would predict an exponent $\beta \approx \tfrac{1}{2}$ in the time dependence $\sigma_D \propto t^{\beta}$. However, this argument makes the critical error of assuming that the rate of heating applied to the disc is directly translated into velocity dispersion. This is false because adiabatic changes to the disc (due to its increase in surface density over time) also effect the evolution of $\sigma_D$ (see section \ref{section:adiabatic}). We will postpone the discussion of heating histories until section \ref{section:heating_histories} after discussing our model of the evolution of the disc.

\subsection{Heating Due to Quantum Wavelets}

In models of wave dark matter, there is an added contribution to the fluctuating substructure due to interference of excited modes which produce time-dependent over-densities known as wavelets. Using results of the distribution of these wavelets given by numerical FDM simulations given by \cite{BECDM}, here we analytically estimate their dynamical heating impact. We assume that the density of the wavelets is a fixed multiple of the local mean density such that,
\begin{equation} \label{multiple_of_background}
M_w = A \left(\frac{\lambda_{\text{char}}}{2} \right)^3 \rho(r),
\end{equation}  
where $\lambda_{\text{char}}$ is the local characteristic scale of the wavelets. Because these wavelets are inherently produced by quantum mechanical interference, they will approximately saturate the uncertainty relations,
\begin{equation}
\lambda_{\text{char}} m_a \sigma_H \approx \hbar,
\end{equation}
where $m_a$ is the mass of the FDM particle or axion. However, the numerical factors in this relationship are important for giving an accurate bound on the particle mass. Therefore, we will provide a more detailed treatment. 
\par
The wavefunction encoding these wavelets can be constructed as follows. We assume that the entire mass of the primary FDM halo (external to the central soliton) is in the form of wavelets such that,
\begin{equation} \label{const}
M_w n = \rho(r).
\end{equation}
Such is the case in simulations of idealized virialized FDM haloes \citep{numerical-Schrodinger}. These wavelets are a result of linear interference patterns of the underlying velocity dispersion of particles $\sigma_H$. Classically, the particles would have an approximately Maxwell-Boltzmann distribution:
\begin{equation}
f(\mathbf{v}) \propto \exp\left(-v^2/(2\sigma_{H}^2/3)\right).
\end{equation}
Quantum mechanically, by the Schr\"{o}dinger-Poisson–Vlasov-Poisson correspondence, the wave function is given by:
\begin{equation} \label{phase_dist}
\psi(\mathbf{x}) \propto \sum_{\mathbf{v}} f(\mathbf{x},\mathbf{v})^{\frac{1}{2}} \exp{\left[i m \mathbf{x}\cdot\mathbf{v}/\hbar + 2\pi i \phi_{{\rm rand},\mathbf{v}} \right]}
\, \mathrm{d}^{3}{\mathbf{v}},
\end{equation}
(equation 27, \cite{Schrodinger-Poisson}) where $\phi_{{\rm rand},\mathbf{v}}\in[0,2\pi)$ is a random phase associated with velocity $\mathbf{v}$.
Such a constructed wavefunction is in agreement with local patches of the wavefunction seen in simulated idealized virialized haloes \citep{BECDM}.
In such a distribution, structure is totally suppressed below the de Broglie wavelength $\lambda_{\rm dB} \equiv \hbar/(m_a \sigma_H)$ by the uncertainty principle.
The characteristic size of these wavelets (determined numerically from the peak of the density power spectrum) is a multiple of the suppression scale:
\begin{equation} \label{characteristic_wavelength}
\lambda_{\text{char}} = \frac{ 2 \pi \hbar}{m_a \sigma_H \sqrt{\frac{3}{2}}}.
\end{equation}
The value of A in equation \eqref{multiple_of_background} can be then determined numerically for the distribution given by equation \eqref{phase_dist} as follows. First, a density field $\rho = |\psi|^2$ in a statistically large-enough box ($> 100$ de Broglie Wavelengths) is constructed according to equation \eqref{phase_dist} with velocity dispersion $\sigma_H$ and average density $\rho = \bar{\rho}$. The number of resulting density peaks per unit volume then gives $n$, which, by the constraint of equation \eqref{const}, determines the average mass $M_w$ of the wavelets. Then, $A$ is deduced from equation \eqref{multiple_of_background} using equation \eqref{characteristic_wavelength} for the characteristic size. Numerically, we find $A\simeq 2.2$ for a wavefunction that encodes an underlying Maxwell-Boltzmann distribution.
\par
The heating due to wavelets is approximately given by,
\begin{subequations} \label{FDMheating}
\begin{align}
\mathcal{H} & = \frac{2 \pi A}{3 \sqrt{3}} \left( \frac{2 \pi \hbar }{m_a} \right)^3 \frac{(\rho(r) G)^2}{\sigma_H^4} \log{\Lambda} ;
\\
\Lambda & = \frac{P \sigma_H}{\lambda_{\text{char}}} = \frac{m_a \sigma_H^2 \sigma_D}{2 \pi^2 \hbar G \Sigma} \sqrt{\tfrac{3}{2}} I_P(z_m / h) .
\end{align}
\end{subequations}
Therefore, using $\Sigma = \Sigma_0 e^{-r/r_0}$ we can express the rate of heating as a function of radius. Under the same assumptions that the considered dynamics are the dominant contribution to disc thickening, as before, the rate of heating given by equation \eqref{FDMheating} depends on $\sigma_D$ only logarithmically. Therefore, the square of the velocity dispersion increases with an approximately constant rate predicting the same $\sigma_D$ -- $t$ relationship as in the case of subhaloes. Once again we note that, since we are considering both components of disc heating as ultimately being reflected in the z-velocity dispersion, we here provide an upper bound on the effect. 

\subsection{Modeling Adiabatic Changes in the Disc} \label{section:adiabatic}

We have computed the rate that energy is delivered to the galactic disc under a number of assumptions about the underlying properties of the dark matter. However, we have yet to consider the somewhat more subtle issue of where that energy ends up. There are a number of discussed effects which might transfer the energy pumped into disc stars by density fluctuations out of the disc such as spiral modes which carry away energy and angular momentum. However, these energy transport effects are, in aggregate, thought to be small and primarily significant for stars on orbits which are resonant with respect to the spiral modes \citep{Baba_spiral_structure, dobbs_baba_2014, Grand_radial_migration, secular_evolution}. Here, we assume that energy loss from the disc itself is negligible. However, even assuming no loss of energy in the disc, the heating delivered does not directly show up as an increase in the velocity dispersion. As the disc changes shape due to heating ``puffing up'' the scale height and additional matter accreting on to the disc, its potential energy changes as well. Some of the heating shows up not in velocity dispersion but in the form of this additional potential energy. Furthermore, as additional matter falls on to the disc, although we have assumed no energy is lost, the total energy does increase. We make the approximation that these changes to the shape and mass of the disc are slow i.e. adiabatic with respect to the motions of individual stars. Since these processes are so slow, this is a very good approximation; the time-scale for vertical oscillations is $P \sim \SI{10}{\mega \year}$ while the time-scale for changes in disc shape and mass are $\sim \SI{10}{\giga \year}$.
\par
Given the adiabatic assumption, the adiabatic invariant action $J_z$ for stars in vertical motion is an (approximate) conserved quantity of the motion in the absence of non-adiabatic heating by impulsive perturbations. Such heating gives flow equations for the actions. Averaging over the motions of all stars in the disc, we find the time-dependence of the overall vertical velocity dispersion to be,
\begin{equation} \label{adiabatic_heating_rate}
\deriv{\sigma_D^2}{t} = \frac{2}{3} \sigma_D^2 \deriv{}{t} \log{\Sigma} + \kappa \EV{\mathcal{H}}, 
\end{equation} 
where $\EV{\mathcal{H}}$ is an averaged (over vertical positions) measure of the heating rate which we calculated earlier in some detail and $\kappa$ is a numerical constant approximately given by $\kappa \approx 0.52$. The constant $\kappa$ reflects the fraction of heating which goes directly into velocity dispersion. Approximating the vertical motion of stars as harmonic, we would expect $\kappa = \frac{1}{2}$ and we find the more detailed computation agrees nicely with the naive prediction. A detailed derivation of this equation is provided in Appendix A. In the case of FDM wavelets, these averaged quantities become,
\begin{subequations}
\begin{align}
\EV{\mathcal{H}} & = \frac{2 \pi A}{3 \sqrt{3}} \left( \frac{2 \pi \hbar}{m_a} \right)^3 \frac{(\rho(r) G)^2}{\sigma_H^4} \log{\Lambda_{\text{avg}}};
\\
\Lambda_{\text{avg}} & = \frac{m_a \sigma_H^2 \sigma_D}{2 \pi^2 \hbar G \Sigma} \sqrt{\tfrac{3}{2}} I_\Lambda.
\end{align}
\end{subequations}
where $I_{\Lambda} \approx 4.81$ is a numerical constant related to the averages over Coulomb logarithms of the vertically distributed stars. 

\subsubsection{The Time Dependence of Surface Density}

\begin{figure}
\includegraphics[width=\columnwidth]{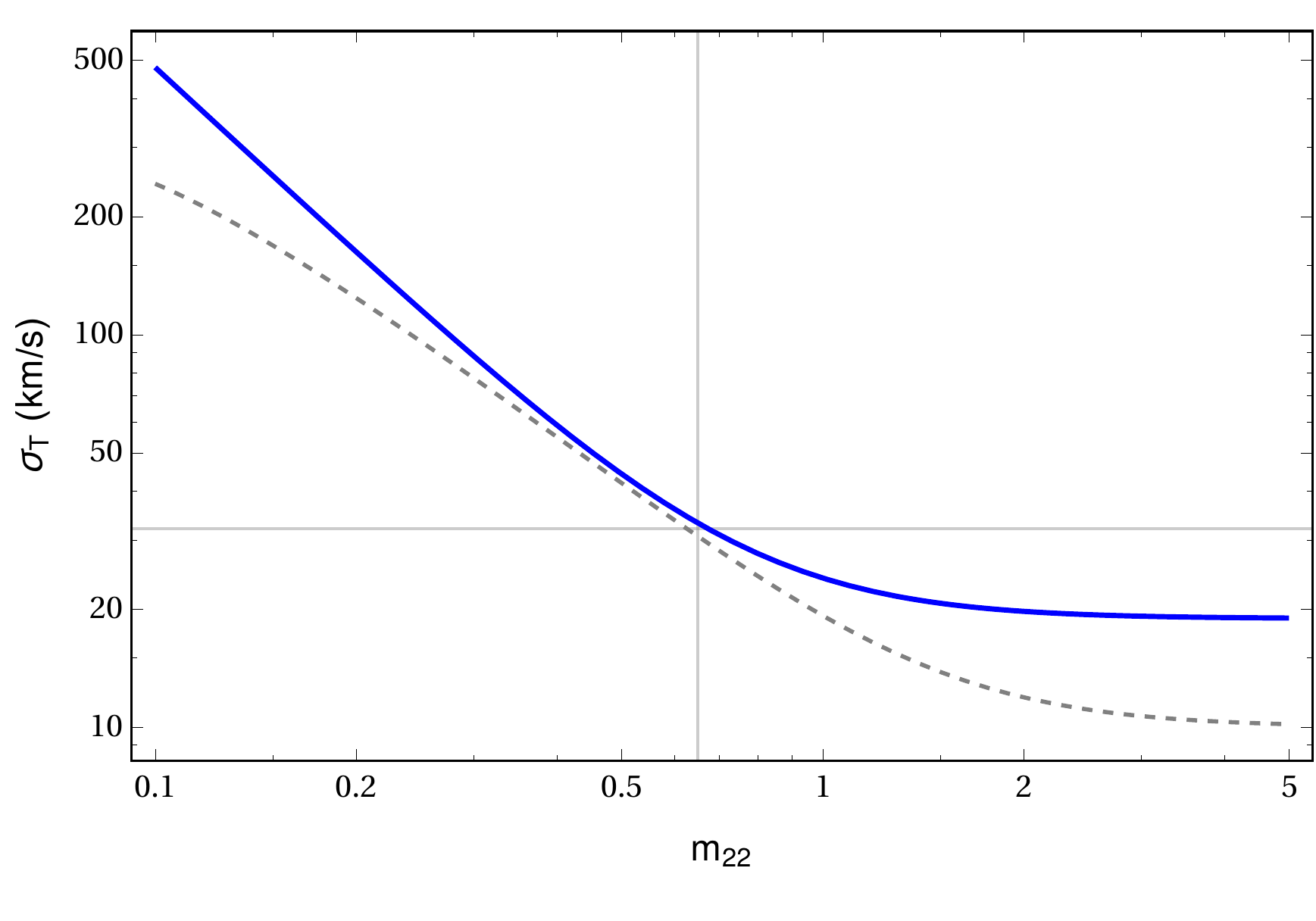}
\vspace*{-5mm}
\caption{Accumulated velocity dispersion at the solar neighborhood $R_{\odot} = \SI{8.3}{\kilo\parsec}$ as a function of axion masses $m_{a}$ in the range $\SIrange{0.1 e-22}{5.0 e-22}{\electronvolt}$ for a Milky Way-like halo. The horizontal line is fixed at $\SI{32}{\kilo\meter\per\second}$, the observational maximum on velocity dispersion of the local Milky Way thick disc. The dashed line represents the accumulated velocity dispersion calculated without taking into account the time dependence of the Coulomb logarithm due to increase in oscillation period with disk velocity dispersion.}
\label{fig:mass_dep_heating}
\end{figure}

To integrate this differential equation \eqref{adiabatic_heating_rate} we must know the explicit time dependence of the disc surface density $\Sigma$. This is estimated by computing the infall rate of baryonic matter on to a density perturbation in a $\Lambda$CDM universe following the methods of \cite{gunn_gott}. The infall rate at high redshift matches the results of \cite{gunn_gott} but is exponentially suppressed in the dark energy dominant epoch. The total accumulated mass is computed numerically, as shown in figure \ref{fig:time_dep}, and the initial conditions at high redshift are chosen by matching the baryonic component to the current mass of the galactic disc. To evaluate such an accretion model we use standard cosmological parameters given by \cite{planck}. Such an analysis predicts a baryonic mass infall rate at the current epoch of $\SIrange{1}{2}{M_{\odot}\year^{-1}}$. Studies of Milky-Way-like spiral galaxies suggest that galactic discs grow ``inside out'' where the inner densities remain approximately constant as the disc grows outwards \citep{Dokkum_in_out}. With this observational evidence in mind, we adopt a simplified model for the evolution of the disc surface density: the central density $\Sigma_0$ remains constant with time and the entire time dependence of the distribution is encoded in a time-dependent scale length $r_s(t)$. This assumption completely constrains the surface density at all radii as a function of time by matching the scale length $r_s(t)$ to the total baryonic mass accreted $M(t)$ at the time $t$. Such a condition gives,
\begin{equation}
\left( \frac{r_s(t)}{r_0} \right)^2 = \frac{M(t)}{M_0}.
\end{equation}  
Therefore, the logarithmic derivative of the surface density which appears as an adiabatic factor in the velocity dispersion time evolution equation is easily computed to be,
\begin{equation}
\deriv{}{t} \log{\Sigma(r,t)} = \frac{1}{2} \left( \frac{r}{r_0} \right) \left( \frac{\dot{M}(t)}{M(t)} \right) \left( \frac{M_0}{M(t)} \right)^{\frac{1}{2}},
\end{equation}
where $M(t)$ is the mass of the disc as a function of time as mass accumulates under gravitational collapse and $M_0$ is the current mass of the disc. Using these forms, the results of integrating equation \eqref{adiabatic_heating_rate} for a Milky-Way-like halo are shown in figures \ref{fig:time_dep} and \ref{fig:mass_dep_heating} (see section \ref{section:application} for details and assumptions of the dark matter model for Milky-Way-like objects). The disc mass (and therefore surface density) increase quickly during early epochs due to rapid accretion in a dense, matter-dominated universe and level off in the current $\Lambda$-dominated cosmology. As show in figure \ref{fig:time_dep}, this accretion profile caused the disc scale height to initially decrease due to the rapid increase in gravitating mass in the disc before entering a period of steady growth from redshift $z \sim 1.5$ -- when heating became dominant -- to the current era.

\section{Applications to a Milky-Way-like Disc and Halo} \label{section:application}

We will now consider, in detail, the distribution of self-gravitating substructure in both CDM and FDM of which the dominant form is an initial distribution of approximately self-similar subhaloes. Additionally, we will consider the peculiar feature of wave dark matter which produces time-dependent density fluctuations on the order of the de Broglie wavelength due to interference. This wavelet substructure provides an additional source of heating whose radial profile is qualitatively very different from the profile of heating due to subhaloes because the wavelets are not self-gravitating objects.

\subsection{Heating in the CDM Paradigm}
\subsubsection{Unresolved Mass Function}

As the primary halo forms, it includes subhaloes and captures additional subhaloes via accretion. The unresolved subhalo mass function refers to the distribution of subhalo masses at the time of accretion before dynamical effects such as tidal stripping have altered the distribution. We assume the unresolved subhalo mass function is very close in shape to the free halo mass function truncated above the primary halo mass. Based on population statistics of subhaloes in high resolution N-body simulations given by \citet{subhalo_abundance} we take this fraction to be $\sim 10\%$. The CDM halo mass function and corresponding subhalo mass function calculated from simulations \citep{pop_of_subhalos, unified_model} are well fit by a power law in the accretion mass $m_{\text{acc}}$ which denotes the mass of a subhalo before modification by the external tidal field. This form matches the free halo mass function, 
\begin{equation}
\frac{\d{n}}{\d{\: \log{m_{\text{acc}}}}} = C \left(\frac{\rho(r)}{M}\right) \left(\frac{m_{\text{acc}}}{M} \right)^{-p} ,
\end{equation}
with $p = 0.9$ and where $M$ is the mass of the primary or host halo.
Furthermore, we fix the total mass in (accreted) haloes as a fraction of the total mass of the primary halo. Therefore,
\begin{subequations}
\begin{align}
M_{\mathrm{haloes}} & = \int m \deriv{n}{(\log{m})} \d{(\log{m})} \d{V} ;
\\
& = C \int \left(\frac{\rho(r)}{M}\right) \left(\frac{m}{M} \right)^{-p} \d{m} \d{V} ;
\\
& = \frac{C}{1-p} \left[ \frac{(f_2 M)^{1-p}}{M^{-p}} \right] = f_1 M ,
\end{align} 
\end{subequations}
so we fix the constant,
\begin{equation}
C = (1 - p)\frac{f_1}{f_2^{1-p}} .
\end{equation}
\subsubsection{Tidal Disruption} 
We adopt a simplistic model of tidal stripping which underestimates the total effect. A tidal radius is calculated by setting the tidal force on a test mass equal to the gravitational attraction of the subhalo. The resulting radius is:
\begin{equation}
R_t = R \left(\frac{m_{\text{acc}}}{2 M(R)}\right)^{1/3} =  R_{\text{max}} \left(\frac{m_{\text{acc}}}{M}\right)^{1/3} f_T(R), 
\end{equation} 
with $f_T(R) \propto R/R_c(\log(1+R/R_c) - R/(R+R_c))^{-1/3}$, where $R_c$ is the core radius of the primary halo. This formula simply comes from the functional form of the enclosed mass as a function of radius for an NFW profile. We then suppose that total truncation occurs outside this radius and no disruption occurs within it. Thus,
\begin{equation}
\frac{T_R(m_{\text{acc}})}{m_{\text{acc}}} = 
\begin{cases}
\frac{\log{(1+R_t/r_c)} - R_t/(r_c+R_t)}{\log{(1+c)} - c/(1+c)},
& \text{if } R_t < r_{\text{max}}
\\
    1,              & \text{otherwise}
\end{cases}
\end{equation} 
However, $m_{\text{acc}} = 200 \rho_0 \frac{4}{3} \pi c(m)^3 r_c^3$ and $r_{\text{max}} = c(m) \: r_{c}$ so we get,
\begin{equation}
R_t = r_{\text{max}} f_T(R).
\end{equation}
Therefore,
\begin{equation} \label{trunc}
\frac{T_R(m_{\text{acc}})}{m_{\text{acc}}} =
\begin{cases}
\frac{\log{(1 + x)} - x/(1 + x)}{\log{(1+c)} - c/(1+c)},& \text{if } f_T(R) < 1
\\
1, & \text{otherwise}
\end{cases}
\end{equation} 
where $x = c(m) \cdot f_T(R)$. Since $c$ is a weak function of $m$, the remaining mass fraction is also a weak function of $m$. Therefore, in agreement with the results of \citet{unified_model}, the final resolved subhalo mass function is approximately also decoupled in mass and radius. High-resolution CDM simulations prominently exhibit the depletion of subhaloes with orbits passing below $r < \SI{20}{\kilo \parsec}$ inside a halo like the Milky Way's. The cosmological high-resolution `zoom-in' simulations of a $M = 10^{12} M_{\odot}$ primary halo performed by \cite{subhalo_truncation_sims} show extreme depletion due to tidal effects of nearly all subhaloes within $\SI{15}{\kilo\parsec}$. However, recent analytical estimates suggest that extreme depletion of subhaloes due to total tidal disruption found in state-of-the-art N-body simulations is, in fact, a numerical artifact \citep{dark_matter_substructure_numerics}. Furthermore, \cite{disruption_numerics} show that the dominant tidal influence on CDM subhaloes is due to the tidal field of the host halo and disc rather than subhalo-subhalo interactions and that physical disruption of CDM substructure due to tidal stripping or shocks is extremely rare. Accordingly, the model we employ produces comparably gentler subhalo mass depletion in the range $10 - \SI{15}{\kilo \parsec}$ and heavily strips mass from all subhaloes within $\SI{10}{\kilo\parsec}$ rather than completely disrupting them. \cite{subhalo_truncation_sims} conclude that the tidal field of the central galaxy is responsible for the severity of subhalo depletion and that the disc potential will primarily outright destroy transiting subhaloes rather than partially strip their mass. Our underestimate of tidal disruption is due to neglecting to accurately account for the sharp tidal field of the galactic disc. However, we find that, within the range where the disc accounts for significant subhalo depletion, heating due to subhalo dynamics is, in any case, unimportant. Therefore, we do not expect the detailed mechanism of tidal depletion to significantly affect our results. 

\subsubsection{Disc Heating Results due to Subhalos}

\begin{figure*}
\includegraphics[width= 17cm]{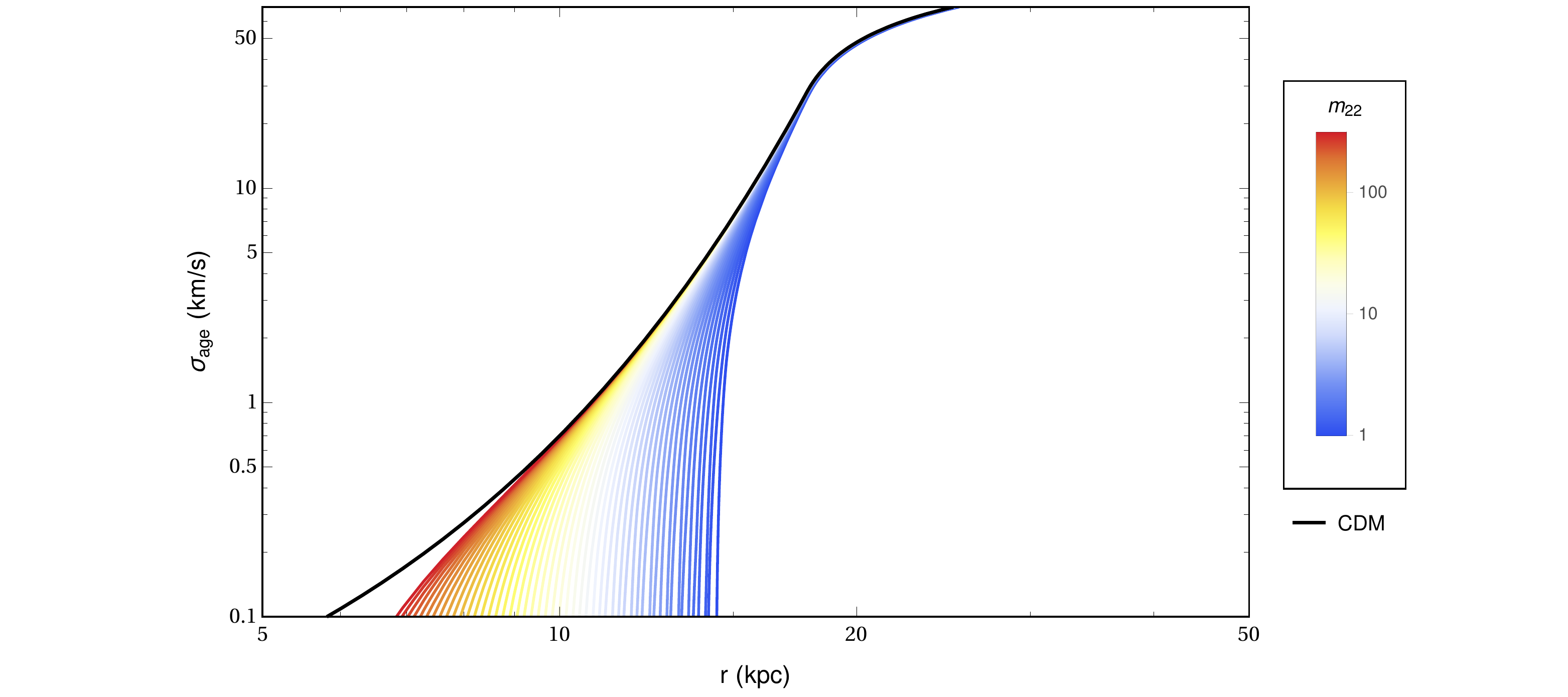}
\caption{Rate of heating due to dynamical subhaloes as a function of radius for a Milky Way-like halo. The heating is calculated for subhaloes in CDM and FDM with a mass range $1 \le m_{22} \le 10^{2.5}$ coloured on a logarithmic scale. Note that FDM subhaloes are less effective at small radii due to a deficiency of small FDM subhaloes below the quantum scale and the greater effectiveness of tidal disruption.}
\label{fig:CDMheating}
\end{figure*}

To give concrete estimation of the effects of various heating mechanisms, we define the quantity,
\begin{align}
\sigma_{\text{age}} &= \left( T_{\text{age}} \cdot \mathcal{H} \right)^{1/2}  
\end{align}
as a standard measure of the approximate heating over the age of a Milky Way-like halo where $T_{\text{age}} \approx \SI{12}{\giga\year}$. 
For a Milky Way-like halo with $M = 0.8 \times 10^{12} \, M_\odot$ and $c = 15$ \citep{milky_way_halo}, we can calculate heating due to subhaloes as a function of radius. We find that the rate of heating is quite weak in the core of the halo where tidal disruption is the dominant effect. However, as seen in figure \ref{fig:CDMheating}, subhalo heating becomes quite efficient near the solar neighborhood $R_{\odot} = \SI{8.3}{\kilo \parsec}$. In fact, at $R_{\odot}$ the rate of heating is $\mathcal{H} = \SI{0.19}{\kilo \meter \squared \per \second \squared \per \giga \year}$ which does accumulate over the lifetime of the galaxy to about 5\% of the observed velocity dispersion of $\SI{32}{\kilo \meter \per \second}$ at $R_{\odot}$. However, somewhat beyond the solar neighborhood at $r = \SI{17}{\kilo \parsec}$ the rate of heating jumps up to $\mathcal{H} = \SI{121.2}{\kilo \meter \squared \per \second \squared \per \giga \year}$ which is marginally in excess of the maximum rate allowed by the total velocity dispersion of the thick disc. However, it is difficult to obtain accurate distribution functions in the outer disc to compare this value to the actual velocity distribution of the thick disc. As the radius increases, heating due to subhaloes becomes far more significant leading to a rapid flaring of the disc and, eventually, destruction of the outer disc (c.f. equation \eqref{fig:disc_scale_height}). As discussed in section \ref{section:disc_destruction}, Subhalo induced heating may be a viable mechanism for limiting the extent of galactic discs.

\subsection{Heating in the FDM Paradigm}

\begin{figure*}
\includegraphics[width=17cm]{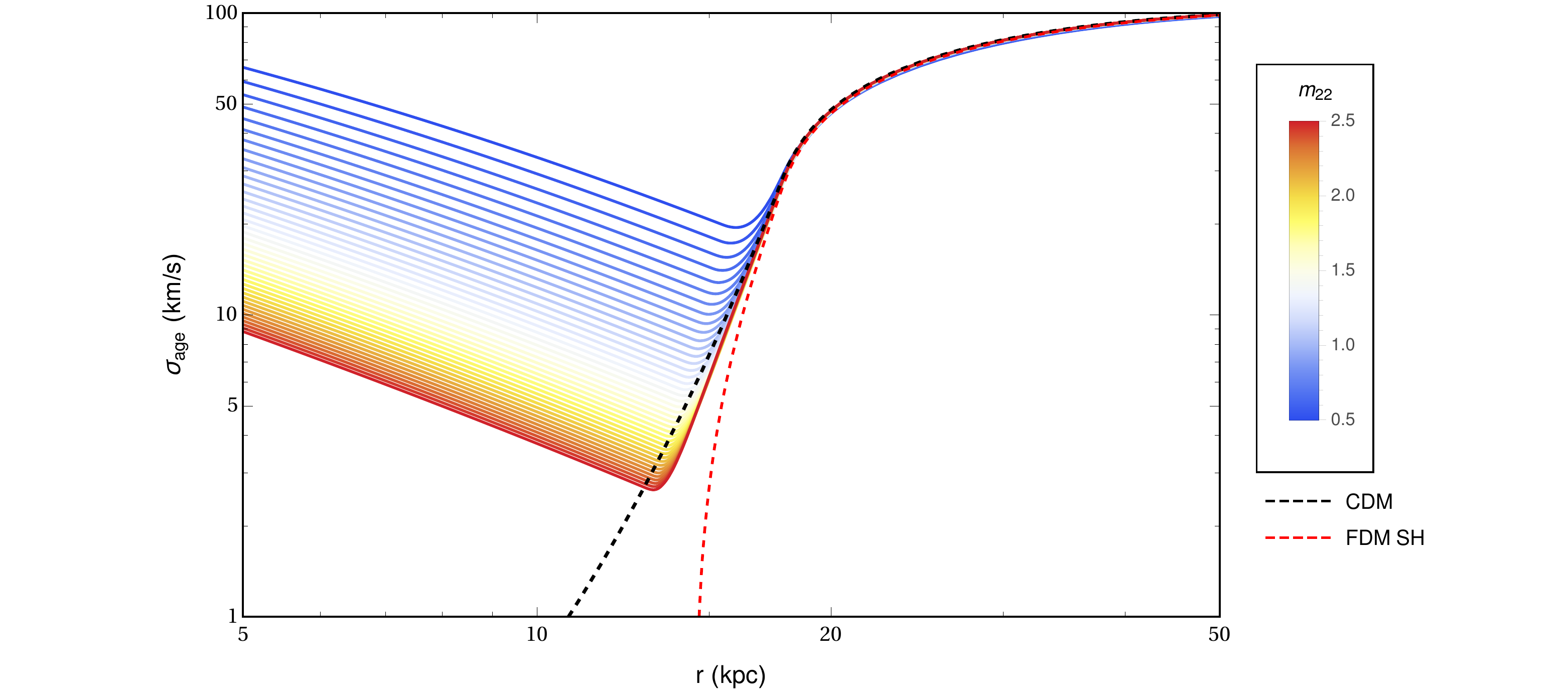}
\caption{Total Rate of heating due to FDM wavelets and subhaloes as a function of radius for axion masses $m_{a}$ in the range $\SIrange{0.25 e-22}{1.75 e-22}{\electronvolt}$ for a Milky Way-like halo. Dashed lines give the heating to subhaloes alone in the CDM and FDM models (with $m_{22} = 1$) respectively.}
\label{fig:radiusheating}
\end{figure*}

\begin{figure}
\includegraphics[width=\columnwidth]{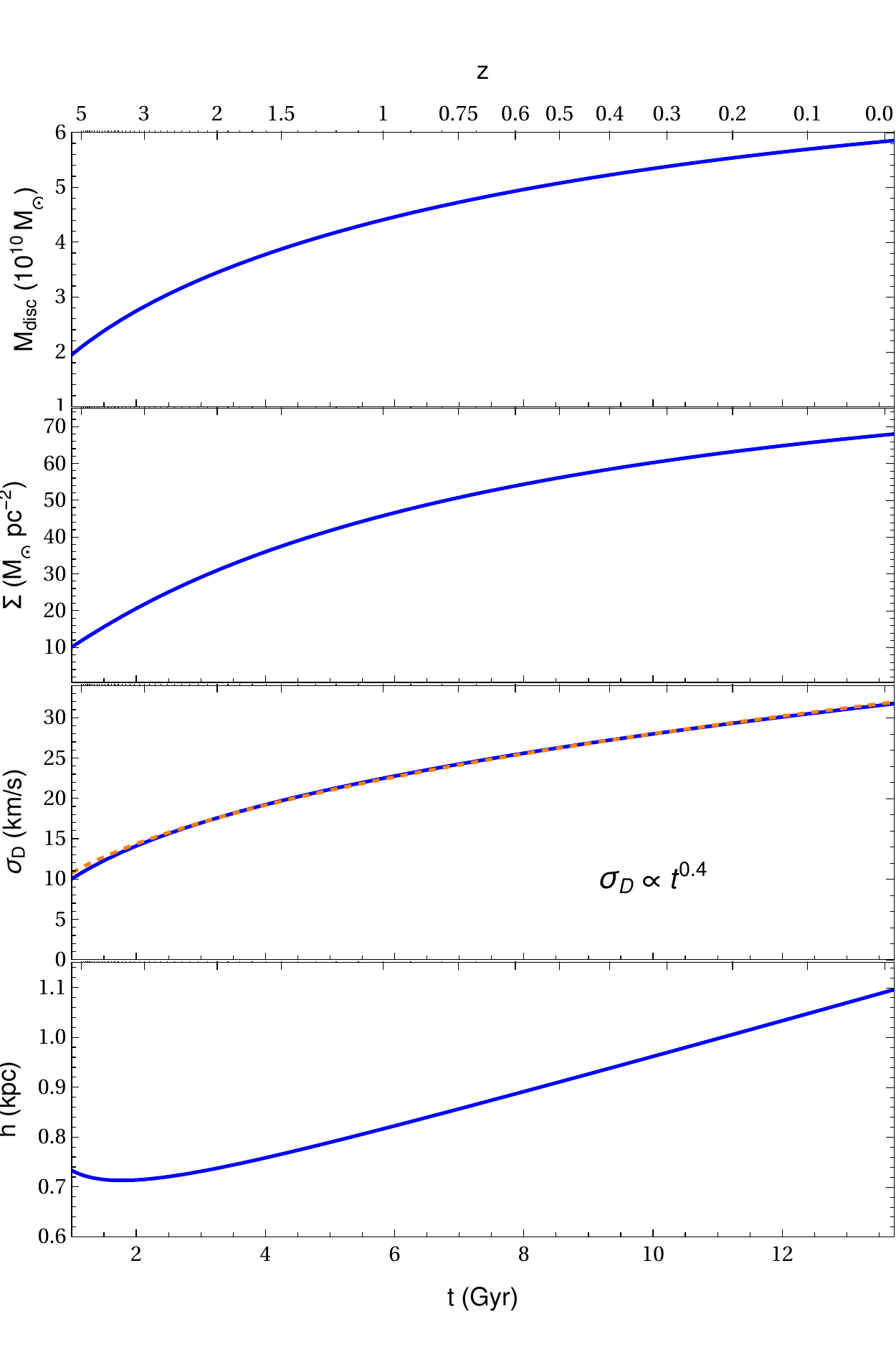}
\vspace*{-5mm}
\caption{Total mass of the disc, local surface density, velocity dispersion, and scale height of the old disc near the solar neighborhood at $r = R_{\odot}$ as a function of cosmic time (and redshift) showing the accretion of baryonic matter due to gravitational collapse. Such an accretion model was evaluated using cosmological parameters $\Omega_{m} = 0.31$, $\Omega_{b} = 0.044$, $\Omega_{\Lambda} = 0.71$, and $H_0 = \SI{68}{\kilo\meter\per\second\per\mega\parsec}$ \citep{planck} and fit to match the current baryonic mass of the Milky Way disc. The surface density and scale height are found by integrating the action equation \eqref{adiabatic_heating_rate} starting from $T_{\text{age}} \approx \SI{12}{\giga\year}$ where heating is produced by FDM density fluctuations with an the axion mass of $m_a = \SI{0.7 e-22}{\electronvolt}$. The dashed orange line shows the best power-law fit for the time evolution of vertical disc velocity dispersion: $\sigma_D \propto t^\beta$ with $\beta = 0.4$.}
\label{fig:time_dep}
\end{figure}

\subsubsection{The Subhalo Mass Function}

The fluctuations due to subhaloes dynamics are reduced in FDM compared to CDM because the halo mass function is suppressed at low-mass due to the quantum Jeans scale. The FDM halo mass function is calculated numerically from the Press--Schechter formalism \citep{substructure_FDM, marsh} and the soliton profile is fitted from simulations by \cite{schive_solitons}. 

\subsubsection{Tidal Disruption}
 
We employ an identical method for truncating subhaloes due to the tidal stripping of the primary halo. However, in FDM, light subhaloes are more easily tidally disrupted via a runaway soliton reformation effect. Since the mass and radius of a soliton are inversely related \citep{solitons}, if the tidal radius of a subhalo lies within its soliton core then mass from the soliton will be stripped away which, due to the self-gravitating quantum condition,
\begin{equation}
R \approx \frac{\hbar^2}{m_a^2 M G},
\end{equation} 
forces the soliton to grow in size and decrease in density causing a runaway effect which completely destroys the subhalo. This runaway process occurs because, as mass is stripped from the soliton, the soliton grows due to quantum pressure (previously halted by gravitational attraction) which pushes more mass outside the tidal radius. To implement this runaway effect, we modify our truncation function by setting $T_R(m) = 0$ if the tidal radius lies within the soliton radius. 
        Furthermore, due to the solitonic cores of haloes in FDM, the remaining mass fraction after tidal disruption is strongly dependent on the subhalo mass because the shape of the soliton profile and behavior are strongly mass-dependent. The decline in heating rate due to subhaloes within \SI{20}{\kilo\parsec} shown in figure \ref{fig:CDMheating} is due to tidal destruction. Note that FDM subhaloes are less effective at small radii due to a deficiency of small FDM subhaloes below the quantum scale and the greater effectiveness of tidal disruption. We can now compare the heating due to subhaloes in the FDM and CDM paradigms and a function of the particle mass. The results are summarized in figures \ref{fig:CDMheating} and  \ref{fig:radiusheating}. Note that subhalo-induced heating is vastly suppressed in FDM with a low particle mass due to the combination of enhanced truncation and the suppression of small-scale power below the quantum scale. Because the enhanced truncation is due to a runaway process, the tidal disruption of FDM subhaloes is highly sensitive to the tidal radius and therefore the position of the subhalo within the primary halo.
This effect is apparent in the steeper decline,  due to tidal destruction, of the heating rate of subhaloes for FDM within \SI{20}{\kilo\parsec} as illustrated in figure \ref{fig:CDMheating}.
 
\subsubsection{Disc Heating Results Due to Wavelets}

Although the fluctuations due to subhalo transits are reduced in the FDM paradigm in comparison to CDM, for a moderately light particle mass ($m_a \approx \SI{e-22}{\electronvolt}$) the dominant effect is due to wavelets, the interference fringes produced by standing FDM waves. In the limit of large particle mass, FDM recreates the results of CDM, since the subhaloes act identically and the power in wavelets is strongly suppressed by a factor of $m_{22}^{-3}$. For very low particle masses, this power grows rapidly and easily exceeds observational bounds for the velocity dispersion of the Milky Way thick disc. We are primarily interested in the intermediate region in which the transition from wavelet dominated fluctuations to subhalo dominated fluctuations occurs. Figure \ref{fig:radiusheating} illustrates that the transition from wavelet dominated heating to subhalo dominated heating occurs approximately at $r \approx \SI{20}{\kilo \parsec}$. For large particle mass, $m_{22} \approx \SI{2.5e-22}{\electronvolt}$, the minimal heating occurs at $r \approx \SI{10}{\kilo \parsec}$. The radius of the minimal heating increases with decreasing particle mass. Furthermore, as the particle mass decreases, the transition between the wavelet dominated and subhalo dominated regimes becomes smoother and the minimum between the two regimes is ameliorated.   

\section{Discussion}

\begin{figure*}
\includegraphics[width=18cm]{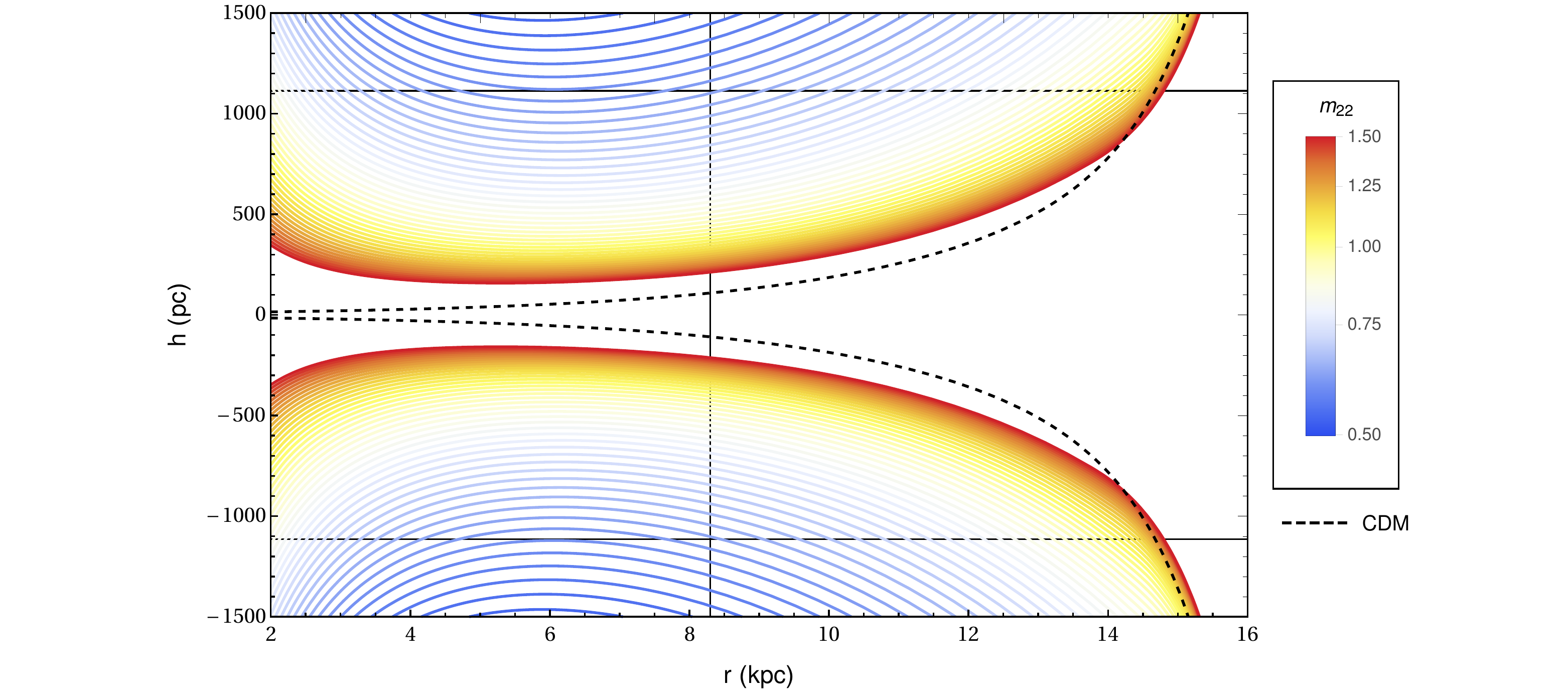}
\vspace*{-5mm}
\caption{Profile of the galactic disc thickness induced by heating from dark matter density fluctuations as a function of radius for a Milky Way-like halo. The vertical velocity dispersion of the old stellar disc is assumed to be entirely produced by heating due to dark matter density fluctuations (both wavelets and subhaloes). The gray lines mark the galactic radius of the solar neighborhood and the scale height of the thick disc at that radius. The inner parts of this figure may not be reliable due to interactions between the disc and the stellar bar which we have not accounted for. However, the enhanced thickening interior to the solar radius is a notable prediction from FDM dynamics which does not appear in CDM calculations.}
\label{fig:disc_shape_FDM}
\end{figure*}

\begin{figure*}
\includegraphics[width=18cm]{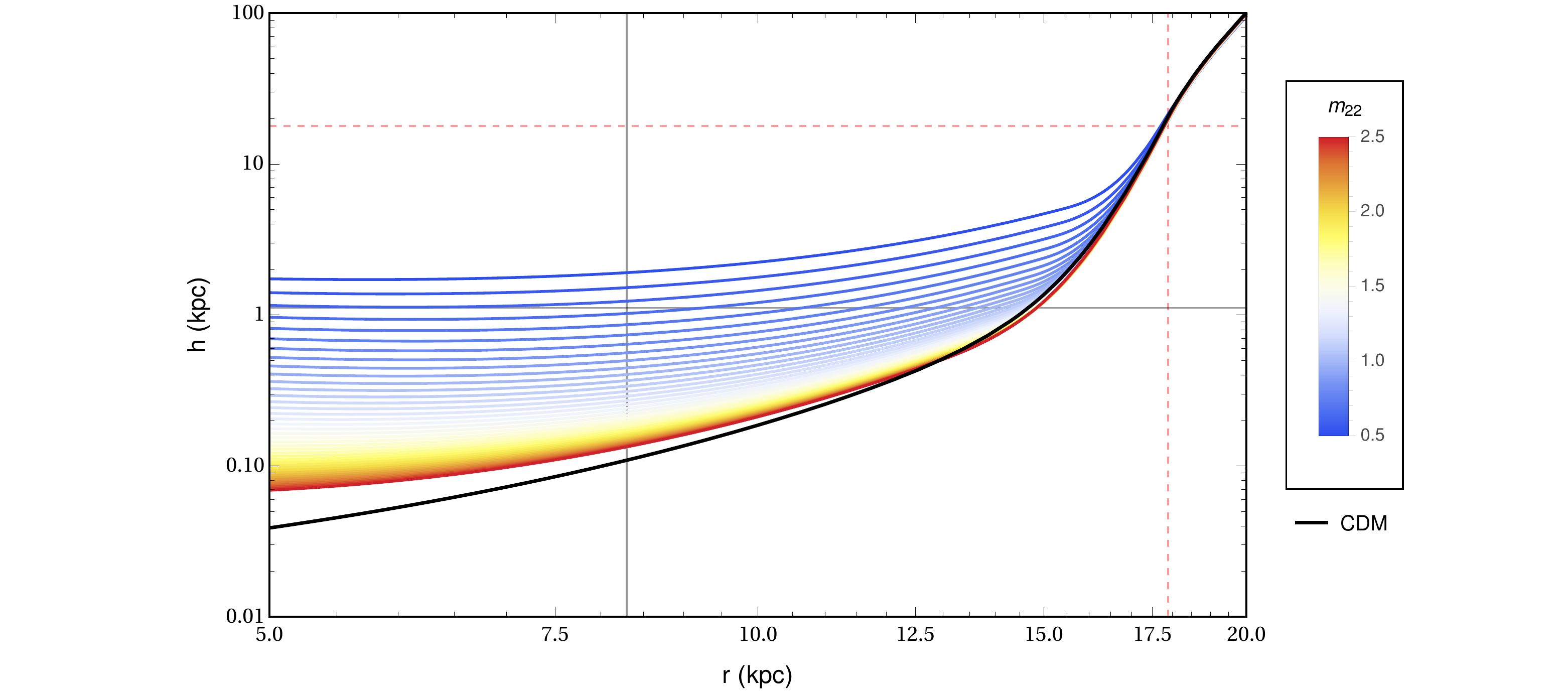}
\vspace*{-5mm}
\caption{Galactic disc scale height induced by heating from dark matter density fluctuations as a function of radius for a Milky Way-like system. The vertical velocity dispersion of the old stellar disc is assumed to be entirely produced by heating due to dark matter density fluctuations (both wavelets and subhaloes). The dashed red lines mark the point of complete disc destruction. The gray lines mark the galactic radius of the solar neighborhood and the scale height of the thick disc at that radius.}
\label{fig:disc_scale_height}
\end{figure*}

\subsection{Disc Destruction at Large Radii} \label{section:disc_destruction}

Given a dark matter paradigm, we have shown how to estimate the heating of the disc caused by dynamical substructure (referring both to subhaloes and, in the case of FDM, to wavelets). If we make the further assumption that dark substructure is the dominant disc thickening effect, then we have a method of predicting the actual velocity dispersion of a disc as a function of radius. However, in a galaxy with a known surface density profile, there is a simple relationship between the velocity dispersion and the scale height of a disc given by equation \eqref{scale}. Therefore, we can compare the disc shapes predicted by CDM and FDM shown in figure \ref{fig:disc_shape_FDM} respectively. Subhaloes alone cause very little thickening out to a radius of approximately $\SI{15}{\kilo \parsec}$ at which point the disc flares exponentially. This rapid growth in thickness will cause the outer disc to become unstable and lead to disc destruction beyond a certain radius. FDM wavelets produce a more leisurely disc flare beginning near $r \approx \SI{10}{\kilo \parsec}$ before transitioning into the subhalo driven exponential flare at $\SI{15}{\kilo \parsec}$. The wavelet induced gentle flare with a slope of roughly $\delta = 0.1$ (that is $\SI{100}{\parsec}$ of scale height per kiloparsec of radius) is consistent with LAMOST survey data which indicate a strong flaring phenomenon of all stellar populations and, in particular, a flaring slope of $\delta = 0.124$ for the thick disc \citep{LAMOST}. 
\par 
We consider a possible criteria to determine the point at which a galactic disc is destroyed by thickening,
\begin{equation}
\frac{h(r)}{r} > 1.
\end{equation}
This clearly gives an upper bound on the endpoint of the disc. Using this criterion, if the disc is thickened by CDM subhaloes alone then the disc will be destroyed past a radius of about $\SI{18}{\kilo \parsec}$. Figure \ref{fig:disc_scale_height} shows that the disc shape is not highly sensitive to particle mass at large radii and therefore to the total rate of heating. Therefore, CDM and FDM paradigms do not significantly differ in predicting disc destruction. In fact, the disc flare is mostly determined by density profile of the disc itself.  
\par
FDM wavelets on the other-hand can produce significant thickening at all radii. For example, near the solar neighborhood $r \approx R_{\odot}$, an axion of mass $m_a = \SI{0.7 e-22}{\electronvolt}$ will lead to a scale height of $\SI{1.1}{\kilo\parsec}$ if both disc heating processes are effective. If only the lower bound given by tidal effects is allowed, then the time scale over which significant disc thickening occurs is much longer than the current age of the universe \citep{ultralight}. Wavelets will also give rise to exponentially flaring discs. However, because wavelets produce their maximum power at small radii and subhaloes at large radii, the disc flaring caused by wavelets is significantly weaker than their subhalo-derived counterpart. Furthermore, the transition point from an approximately flat disc to a flaring disc is highly sensitive to particle mass. Therefore, the radius at which disc destruction takes place is quite sensitive to $m_{a}$ although it will take place at a larger radii than the comparable CDM prediction because there are fewer low-mass subhaloes predicted by FDM than by CDM.

\subsection{Constraints on Particle Mass}

\begin{figure}
\includegraphics[width=\columnwidth]{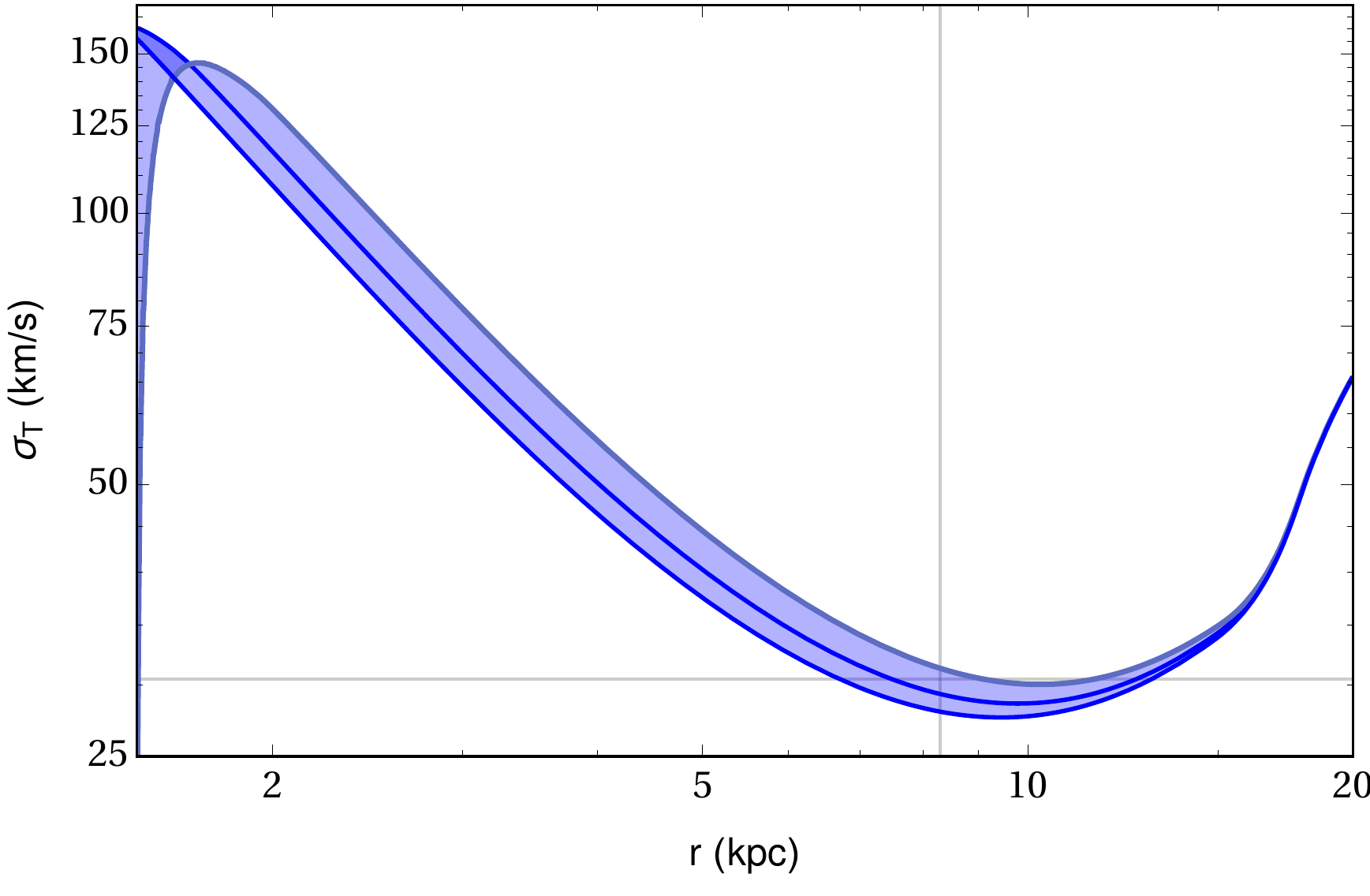}
\vspace*{-5mm}
\caption{Vertical velocity dispersion profile of the disc at the current epoch as a function of radius where heating is produced entirely via FDM density fluctuations (including wavelets and subhaloes) for the most probable particle mass of $m_{a} = \SI{0.7 e-22}{\electronvolt}$ necessary to entirely explain the observed thick disc locally. Error bars are given at the $2 \sigma$ level. Light gray lines indicate the solar neighborhood and the observed velocity dispersion of the disc at that radius respectively.}
\label{fig:heating_shape}
\end{figure}

The Milky Way disc has two components, a recently formed thin disc with low velocity dispersion and an old thick disc with relatively large velocity dispersion. The thick disc has locally a velocity dispersion of approximately $\SI{32}{\kilo\meter\per\second}$ in its thickest parts \citep{milky_way}. Therefore, the velocity dispersion caused by all FDM dynamical heating cannot exceed this value. We find that CDM does not produce sufficient heating to be constrained by observational measurements of the local  Milky Way disc. However, the heating due to wavelets in FDM is strongly dependent on particle mass and far more efficient than other  forms of dynamical density fluctuations. In the low-mass limit, the heating may vastly exceed observational bounds. Using the results quoted for the velocity dispersion of the thick disc as a strict cutoff, we derive a lower bound at $2 \sigma$ confidence (see section \ref{section:errors}) on the mass of the FDM particle,
\begin{equation}
m_a > 0.6 \times \SI{e-22}{\electronvolt}
\end{equation}
under the assumption that all disc heating is reflected in the z-velocity dispersion. If we further assume that dynamical FDM substructure is the primary source of disk heating then the vertical velocity dispersion in the solar neighborhood constrains the entire heating profile at all radii. For the most probable value of $m_{a}$, the maximum heating occurs at $r = \SI{1.7}{\kilo \parsec}$ with a value of $\sigma_{T} = \SI{124}{\kilo\meter\per\second}$. The entire heating profile for this value of the FDM particle mass is shown in figure \ref{fig:heating_shape}. We hope that detailed studies of the Milky Way disk stars outside the solar neighborhood made possible by recent Gaia data will test the validity of this proposed heating curve.   

\subsection{Heating History}
\label{section:heating_histories}

A separate observational test of our theoretical treatment derives from our prediction for the heating history of the disc. If heating is dominant in driving the evolution of $\sigma_D$ then, since we predict a heating rate that depends only logarithmically on $\sigma_D$, a naive computation would conclude that both dark matter paradigms predict an approximately constant increase in the velocity dispersion of the disc and therefore a time dependent velocity dispersion of $\sigma_D \approx t^{\beta}$ with $\beta = \tfrac{1}{2}$. This disagrees with analyses which suggest that the Milky Way thick disc has increased in velocity dispersion with exponent $\beta \approx \tfrac{1}{3}$ \citep{heating_history}. However, these arguments both assume that dark substructure alone (ignoring, for example, the contribution due to spiral structure heating effective at low values of $\sigma_D$) is responsible for heating the disc and, critically, fail to include adiabatic effects. If we additionally include the effects of, for instance, scattering from GMCs which gives an evolutionary heating trajectory with $\beta = \frac{1}{4}$, these multiple sources of heating may combine to produce a lower effective exponent. Furthermore, such an analysis ignores the adiabatic changes to the velocity dispersion; when the heating history shown in figure \ref{fig:time_dep} is fit to a power-law, we find an exponent of $\beta = 0.4$ closer to the observed $\beta \approx \frac{1}{3}$ (see figure \ref{fig:time_dep}). The recent findings of \cite{Gaia_vertical_motions} draw upon APOGEE and Gaia data to estimate vertical heating as a function of age.  In the solar neighborhood, they find that the mean vertical action of local populations of disc stars scales with age as $\widehat{J}_z \propto t^{\gamma}$ with $\gamma \approx 1$ which, in the harmonic potential approximation, is approximately a heating history of $\sigma_z \propto t^\beta$ with $\beta \approx \frac{1}{2}$. However, applying our analysis including anharmonicities and the effects of adiabatic changes (see Appendix A) predicts, for the measured action -- age relation $\hat{J}_z \propto t^{\gamma}$ with $\gamma \approx 1$, velocity dispersion scaling with $\beta \approx 0.4$ in agreement with our theoretical prediction for FDM heating. Superficially, this result is inconsistent with heating due primarily to GMC scattering which predicts $\beta = \frac{1}{4}$. However, $\widehat{J}_z$ is not exactly the evolutionary heating path but reflects both the action of stars at birth and their subsequent heating history. \cite{Gaia_vertical_motions} argue GMC scattering can be salvaged to be consistent with $\gamma \approx 1$ if the scattering amplitudes and heating efficacy are functions of time in a certain physically reasonable way. However, when the adiabatic effects of the disc are fully incorporated, heating due to dynamical substructure automatically predicts a heating history of the thin disc consistent with the observed value $\gamma \approx 1$ and $\beta \approx \frac{1}{3}$ without the need for time-dependent parameters as well as explaining the existence of the thick disc which GMC scattering does not. In fact, the heating we predict from FDM density fluctuations naturally explains the phenomenon of increasing vertical action with radius in the range 10-\SI{15}{\kilo \parsec}.
That said, our analysis fails to explain the gradual observed increase in $\gamma$ as a function of radius and predicts a greater degree of heating to the interior of the disc within $R_{\odot}$ than is found by \cite{Gaia_vertical_motions}. If both FDM substructure and GMC scattering operate in line with the current understanding, it is plausible, if not likely, that a combination of these effects best explains the heating data.

\subsection{Comparisons With Other Current Work}

The contemporary study by \cite{stellar_streams_bound} has also determined a lower bound on the mass of the FDM axion using similar methodology to study the thickening of thin stellar streams by FDM wavelets. These calculations exhibit the same scaling in terms of $m_{a}$ and similar heating versus time profiles. Thin stellar streams give a stricter lower bound of $m_a > \SI{1.5 e-22}{\electronvolt}$ which is higher than our bound derived from an analysis of disc thickness. While an axion mass in excess of $\SI{1.5 e-22}{\electronvolt}$ is insufficient to explain the thick disc alone, it is worth noting that galactic processes from multiple regimes suggest FDM models with particle masses within small numerical factors.    
\par 
\cite{relaxation} recently considered relaxation processes in the FDM paradigm by studying diffusion in stochastic density fields. Of many results, the authors calculate an effective heating time-scale for FDM fluctuations to pump energy into massive bodies of
\begin{equation}
T_{\text{heat}} = \frac{3 m_a^3 \sigma^6}{16 \pi^2 G^2 \rho^2 \hbar^2 \log{\Lambda_{\text{FDM}}}} .
\end{equation}  
This effective heating time-scale exhibits the same scaling relations and is comparable to the ratio we, from our own analysis, find between the heating rate and the velocity dispersion of the halo. Explicitly, ignoring the time-dependence of $\Sigma$,
\begin{equation}
\left( \frac{1}{\sigma_H^2} \deriv{\sigma_D^2}{t} \right)^{-1} = \frac{3 \sqrt{3} m_a^3 \sigma_H^6}{16 \pi^4 A \kappa G^2 \rho^2 \hbar^2 \log{\Lambda_{\text{FDM}}}} = \frac{\sqrt{3} }{A \kappa \pi^2} T_{\text{heat}}.
\end{equation}
Furthermore, \cite{relaxation} explicitly calculate the effective time-scale for the heating of a spherical population of stars in equation (103). The authors find that, for a singular isothermal spherical halo with parameters comparable to the halo of the Milky Way, FDM heating is only significant at radii less than $\sim \SI{1}{\kilo \parsec}$. This, however, does not contradict the results discussed in this paper, primarily because $T_{\text{heat}}$ gives the time-scale over which FDM fluctuations pump enough energy into massive bodies for their velocity dispersion to become comparable to the velocity dispersion \textit{of the FDM halo}. This far exceeds the required heating to account for the thickening of a galactic disc because the vertical velocity dispersion of the disc is about an order of magnitude smaller than that of the halo. Furthermore, the singular isothermal sphere is not a good model of the density profile of an FDM halo near the solar radius. Since $R_{\odot}$ is within the core radius of the primary halo for reasonable estimates of the NFW halo of the Milky Way, we expect that the density profile goes as $r^{-1}$ rather than $r^{-2}$ as predicted by a singular isothermal sphere model. Essentially, this discrepancy is due to the fact that, near the solar neighborhood, the mass of the Milky Way disc accounts for a significant fraction of the orbital velocity and thus the halo alone does not produce a flat rotation curve (which would imply a singular isothermal sphere).

\par

Therefore, the density, the size of FDM density fluctuations, and thus the rate of heating is a less extreme function of radius in a NFW halo than is the model presented by \cite{relaxation}. Near the solar neighborhood, we would expect the actual heating time-scale to depend on $r^2$ rather than $r^4$ (c.f. equation (103), \cite{relaxation}). These considerations lead to a modified formula for the heating radius,
\begin{multline}
r_{\text{heat}} = \SI{9.28}{\kilo\parsec} \left( \log{\Lambda} \frac{T_{\text{age}}}{\SI{10}{\giga\year}} \right)^{1/2} \left( \frac{v_c}{\SI{200}{\kilo\meter\per\second}} \right)^{-1} 
\\
 \cdot \left( \frac{m_a}{\SI{0.5 e-22}{\electronvolt}} \right)^{-3/2},
\end{multline}
which is the radius such that heating is effective whenever $r_{\star} < r_{\text{heat}}$. Therefore, heating can be effective near the solar neighborhood when $m_{22} \sim 0.5$. Taking into account the shorter time-scale needed to heat the disc up to its current thickness and the less extreme radial falloff in the strength of this heating, the results of \cite{relaxation} are, in fact, consistent with the possibility that FDM fluctuations with a particle mass $m_a = \SI{0.7 e-22}{\electronvolt}$ are the cause of the observed thick disc near the solar radius.

\subsection{Sources of Error}
\label{section:errors}

The confidence interval associated to our bounds on the FDM particle mass derives from estimates of the errors associated with measured parameters of the Milky Way system. Such uncertainties are on the order of $\sim 10 \%$ of their measured values. We estimate that the mass of the FDM particle must be $m_a = \SI{0.68 \pm 0.05 e-22}{\electronvolt}$ where the dominant source or error derives from the uncertainty in the measured local vertical velocity dispersion of $\sigma_D = \SI{32 \pm 1}{\kilo\meter\per\second}$ \citep{milky_way} which provides our bound on the total heating.

\par

There is one significant effect that we know has been neglected in our calculation. In computing its effects on heating the disc stars, we treated the distribution of FDM  dark matter as if it were unaffected by the baryon distribution embedded within it. In fact the baryons
dominate in the inner parts of the galaxy and will pull in the FDM component, increasing its density and the dynamical effects produced by wavelets. An estimate of the density increase \citep{mass_profile_milky_way} is approximately $40 \%$ leading to a corresponding underestimate of the heating by that amount. Given that the heating rate scales with the FDM particle mass as $m_a^{-3}$, this increase in heating corresponds to raising our lower bound to the particle mass by $12 \%$ to $m_a > \SI{0.71 \pm 0.05 e-22}{\electronvolt}$. Since a full calculation which includes the effect of baryons ``pulling in'' dark matter produces a stricter lower bound on $m_a$, our calculation neglecting this effect still produces a valid, although weaker, lower bound. 
The heating contribution of self gravitating subhaloes for either FDM or CDM is negligible since these dominate in the outer parts of the system which is much less affected by the baryonic component.

\section{Conclusion}

We find that the heating of Milky-Way-type old stellar discs by dark gravitational fluctuations will lead to  radius-dependent flaring and disc thickness in both CDM and FDM paradigms. In addition, FDM quantum wavelets can be effective at heating the inner disc regions accounting for the thick disc and the observed flaring of star populations. Furthermore, we have demonstrated that heating from dark substructure in the extremities of the disc, predominantly due to subhalo transits, can cause complete destruction of the outer disc at radii greater than \SIrange{15}{20}{\kilo\parsec} in both CDM and FDM paradigms and provides a possible explanation of the radial limits of galactic discs. A particle mass in excess of $m_a > \SI{0.6 e-22}{\electronvolt}$ is required to avoid exceeding observational bounds on heating of thick disc stars in the Milky Way. Correspondingly, an FDM scenario with particle mass $m_a \approx \SI{0.7 e-22}{\electronvolt}$ can, in fact, explain the observed thick disc and power-law increase in velocity dispersion as a function of stellar age with $\sigma_D \propto t^{0.4}$.

\par

It is natural to consider how FDM and CDM substructure may influence the discs of galaxies besides the Milky Way. In general, this is a very difficult problem because accurate measurements of the haloes, discs, and vertical velocity dispersion of other galaxies are quite difficult to obtain. Since our model of dynamical substructure will produce vertical heating to any disc embedded in a dark matter halo, the family of superthin galaxies are a particularly interesting test case. The recent study of the superthin galaxy FGC1540 by \cite{superthin} estimates properties of the galaxy's dark matter halo in enough detail to determine whether the thickness (or lack there of) of FGC1540 is inconsistent with our model.  
Without presenting the details, we can say that the system in question has a very young very thin disc and an older thick disc, with the latter not inconsistent with FDM for particle masses on the order of $\SI{e-22}{\electronvolt}$ and with the constraint derived from the Milky Way thick disc. The thin disc of FGC1540, on the other hand, must be very young, $\sim \SI{2}{\giga \year}$, for FDM heating with $m_{a} \sim \SI{e-22}{\electronvolt}$ to be consistent with its observed thickness.

\par

Future work should apply these heating models to a variety of disc galaxies whose disc profiles are observationally known to provide a stronger lower bound on the FDM axion mass. Such work might also develop sophisticated modeling for the disruption of outer discs and compare the predictions due to subhalo heating to the observed sizes of nearby galactic discs. Finally, the recent Gaia data provide an opportunity to accurately examine the entire radial profile of vertical stellar velocity dispersion. A study comparing these observational findings to the profiles predicted in this work would be an excellent test of the hypothesis that the thick disc is produced predominantly by dark matter perturbations.   

\section*{Acknowledgments}

We thank Mihir Kulakarni for his code to calculate the FDM halo mass function and his invaluable guidance on this project. We further thank Professor Hsi-Yu Schive for helpful discussions on the nature of FDM solitons and wavelets and Professor James Binney for critique and thoughts on the nature and origin of the old disc. Many thanks to Professor Scott Tremaine for his helpful criticism and insights regarding the transfer of energy to disc thickening from bending modes of the disc versus tidal strain. Support (PM) for this work was provided by NASA through the Einstein Postdoctoral Fellowship grant no. PF7-180164 awarded by the Chandra X-ray Center which is operated by the Smithsonian Astrophysical Observatory for NASA under contract NAS8-03060.

\section*{Appendix A: Derivation of Adiabatic Equation}

Consider the density distribution,
\begin{equation} 
\rho(r, z, t) = \frac{\Sigma(r,t)}{2 h(r, t)} \sech^2{(z/h(r,t))}, 
\end{equation}
with scale height,
\begin{equation}
h = \frac{\sigma_D^2}{\pi G \Sigma},
\end{equation}
such that $\rho$ satisfies the vertical Jeans equation,
\begin{equation}
\frac{1}{\rho} \pderiv{}{z} \left( \frac{1}{\rho} \pderiv{}{z} \rho \right) = - \frac{4 \pi G}{\sigma_D^2}.
\end{equation}
By Gauss's law, the gravitational acceleration is,
\begin{equation}
\mathbf{g}(z) = -4 \pi G \int_0^z \rho(z') \d{z'} = -2 \pi G \Sigma \tanh{(z / h)}, 
\end{equation}
and therefore, the gravitational potential is,
\begin{equation}
\phi(z) = - \int_0^z \mathbf{g}(z') \d{z'} = 2 \pi G \Sigma h \log{\cosh{(z / h)}}.
\end{equation}
Thus, we define the vertical action\footnote{The conventions used here differ slightly from those of \cite{binney_tremaine_2008} and others which define $J = \frac{1}{2\pi} \oint p \d{q}$. We have chosen our action-angle variables such that $\pderiv{J}{E} = P$ while other texts chose their conventions such that $\pderiv{E}{J} = \omega = \frac{2 \pi}{P}$.} as,
\begin{subequations}
\begin{align}
\frac{J_z}{2m} & = \oint p \d{q} = \int_{-z_m}^{z_m} \sqrt{2(\phi(z_m) - \phi(z'))} \d{z'} 
\\
& = \sqrt{4 \pi G \Sigma h^3} \int_{-z_m}^{z_m} \frac{\d{z'}}{h}  \sqrt{\log{\frac{\cosh{(z_m / h)}}{\cosh{(z'/ h)}}}}
\\
& = \frac{2 \sigma_D^3}{\pi G \Sigma} I(z_m / h);
\\
I(u) & = \int_{-u}^u \sqrt{\log{\cosh{u}} - \log{\cosh{x}}} \d{x}.
\end{align}
\end{subequations}
Let $E$ denote the value of the Hamiltonian for the vertical motion of the star in question. Explicitly, 
\begin{equation}
E = H(p_z, z) = \frac{p_z^2}{2 m} + m \phi(z) = m \phi(z_{\text{max}}).  
\end{equation}
We find the period of the orbit from differentiating the action variable with respect to the fixed value of the Hamiltonian along the phase space trajectory defining the action. Thus, the period equals,
\begin{align}
\pderiv{J_z}{E} & = 2 \int_{-z_m}^{z_m} \frac{1}{\sqrt{2 ( \phi(z_m) - \phi(z') )}} \d{z'} 
\\
& = 2 \sqrt{\frac{h}{4 \pi G \Sigma }} \int_{-z_m}^{z_m}  \frac{\d{z'}}{h} \left( \log{\frac{\cosh{(z_m / h)}}{\cosh{(z'/ h)}}} \right)^{-\frac{1}{2}}
\\
& = \frac{2 \sigma_D}{2 \pi G \Sigma} I_P(z_m / h);
\\
I_P(u) & = \int_{-u}^u \frac{\d{x}}{\sqrt{\log{\cosh{u}} - \log{\cosh{x}}}}.
\end{align}
Now we have heating due to an instantaneous velocity impulse of the form,
\begin{align*}
\deriv{v^2}{t} = Q = M \log{(P / \tau)},
\end{align*}  
where $M$ and $\tau$ depend only on the properties of the halo and is independent of the star in question. The constant $\tau$ is a characteristic time-scale for the interactions so the Coulomb integral is $\Lambda = P / \tau$. The change in action of a single star is given by,
\begin{equation}
\deriv{J_z}{t} = \pderiv{J_z}{E} \deriv{E}{t} + \pderiv{J_z}{\lambda_i} \deriv{\lambda_i}{t},
\end{equation}
where the $\lambda_i$ are the parameters of the potential. Furthermore,
\begin{equation}
\deriv{E}{t} = \pderiv{E}{\lambda_i} \deriv{\lambda_i}{t} + \frac{1}{2} m Q,
\end{equation}
and 
\begin{equation}
\pderiv{J_z}{E} = P,
\end{equation}
so putting everything together,
\begin{equation}
 \deriv{J_z}{t} = \left( P \pderiv{E}{\lambda_i} + \pderiv{J_z}{\lambda_i} \right) \dot{\lambda}_i + \frac{mP}{2} Q.
\end{equation}
However, because $J_z$ is an action adiabatic invariant,
\begin{equation}
\pderiv{J_z}{\lambda_i} = - P \EV{\pderiv{E}{\lambda_i}}_{\text{orbit}}.
\end{equation}
Thus if $\lambda_i$ do not change much over the course of an orbit then,
\begin{equation}
P \pderiv{E}{\lambda_i} + \pderiv{J_z}{\lambda_i}  =  P \pderiv{E}{\lambda_i} - P \EV{\pderiv{E}{\lambda_i}}_{\text{orbit}}  \approx 0,
\end{equation}
and therefore,
\begin{equation}
\deriv{J_z}{t} = \frac{m P}{2} Q.
\end{equation}
Plugging in,
\begin{align}
\deriv{}{t} & \left( \frac{2 \sigma_D^3}{\pi G \Sigma} I(z_m / h) \right) 
\\
& = \frac{1}{2} \left( \frac{\sigma_D I_P(z_m / h)}{2 \pi G \Sigma} \right) M \log{\left( \frac{\sigma_D I_P(z_m / h)}{\pi G \Sigma \tau}  \right)}, \nonumber
\end{align}
rearranging,
\begin{align}
\frac{\Sigma} {\sigma_D} \deriv{}{t} & \left( \frac{\sigma_D^3}{\Sigma} I(z_m / h) \right) 
\\
& = \frac{1}{8}  I_P(z / h) M \log{\left( \frac{\sigma_D I_P(z_m / h)}{\pi G \Sigma \tau} \right)}, \nonumber
\end{align}
and then expanding this expression,
\begin{align}
I(z_m / h) & \left( \frac{3}{2} \deriv{\sigma_D^2}{t} - \sigma_D^2 \deriv{}{t} \log{\Sigma} + \sigma_D^2 \deriv{}{t} \log{I(z_m / h)} \right) 
\\
& = \frac{1}{8}  I_P(z_m / h) M \log{\left( \frac{\sigma_D I_P(z_m / h)}{\pi G \Sigma \tau} \right)}. \nonumber
\end{align}
However, this is the action of a single star and is thus $z_m$ dependent. We really care about the average action. So we average all these quantities over $z_m / h$,
\begin{align}
\EV{I(z_m / h)}_{z_m} & \left( \frac{3}{2} \deriv{\sigma_D^2}{t} - \sigma_D^2 \deriv{}{t} \log{\Sigma} \right) + \sigma_D^2 \EV{\deriv{}{t} I(z_m / h) }_{z_m}  
\\
&= \frac{1}{8} M \EV{ I_P(z / h) \log{\left( \frac{\sigma_D I_P(z_m / h)}{\pi G \Sigma \tau} \right)} }_{z_m}. \nonumber
\end{align}
If we assume that there is sufficient thermodynamic coupling to maintain the disk component we are studying in thermal equilibrium, then $x = z_m / h$ is distributed according to $\sech^2{x}$ for all times. Thus,
\begin{align}
\EV{I(z_m / h)}_{z_m} & = \int_0^\infty I(x) \sech^2{x} \d{x} = I_1 \approx 0.759;
\\
\EV{\deriv{}{t} I(z_m / h) }_{z_m} & = \frac{1}{N} \sum_{i = 1}^N \deriv{}{t} I(z_i / h)
\\
& = \deriv{}{t} \frac{1}{N} \sum_{i = 1}^N \deriv{}{t} I(z_i / h) \nonumber
\\
& = \deriv{}{t} \EV{I(z_m / h)}_{z_m} = 0, \nonumber
\end{align}
because the expectation value $\EV{I(z_m / h)}_{z_m}$ is constant. Furthermore,
\begin{align}
& \EV{ I_P(z_m / h)  \log{\left( \frac{\sigma_D I_P(z_m / h)}{\pi G \Sigma \tau} \right)} }_{z_m}
\\
& = \EV{I_P(z_m / h)}_{z_m} \log{\left( \frac{\sigma_D}{\pi G \Sigma \tau} \right)}  + \EV{I_P(z_m/h) \log{I_P(z_m / h)}}_{z_m} \nonumber
\\
& = I_2 \log{\left( \frac{\sigma_D}{\pi G \Sigma \tau} \right)} + I_2 \log{I_{\Lambda}} = I_2 \log{\left( \frac{\sigma_D I_{\Lambda}}{\pi G \Sigma \tau} \right)}, \nonumber
\end{align}
where
\begin{align}
I_2 & = \EV{I_P(z_m / h)}_{z_m} = \int_0^\infty I_P(x) \sech^2{x} \d{x} \approx 4.79;
\\
I_{\Lambda} & = \exp{\left( \frac{1}{I_2} \EV{I_P(z_m / h) \log{I_P(z_m / h)} }_{z_m} \right)} \nonumber
\\
& = \exp{\left( \frac{1}{I_2} \int_0^\infty I(x) \log{I(x)} \sech^2{x} \d{x} \right)} \approx 4.81.
\end{align}
Therefore, plugging back into the heating equation,
\begin{align}
\frac{3}{2} \deriv{\sigma_D^2}{t} = \sigma_D^2 \deriv{}{t} \log{\Sigma} + \frac{I_2}{8 I_1} M \log{\left( \frac{\sigma_D I_{\Lambda}}{\pi G \Sigma \tau} \right)}
\end{align}
We can define some new quantities,
\begin{align*}
I_r & = \frac{I_2}{I_1} \approx 6.31
\\
\kappa & = \frac{I_r}{12} \approx 0.526
\end{align*}
Then the heating equation becomes,
\begin{equation}
\deriv{\sigma_D^2}{t} = \frac{2\sigma_D^2}{3} \deriv{}{t} \log{\Sigma} + \kappa M \log{\left( \frac{\sigma_D I_{\Lambda}}{\pi G \Sigma \tau} \right)}
\end{equation}
This is almost the same as what we might naively expect,
\[ \deriv{\sigma_D^2}{t} = M \log{(P / \tau)} \]
except with three alterations. First, $P$ is replaced by some sort of average period over all possible $z_m$ which makes physical sense since our heating formula should not depend on the $z_m$ of a particular star. Furthermore, we get an extra term,
\[ \frac{2\sigma_D^2}{3} \deriv{}{t} \log{\Sigma} \]
which is the extra heating due to mass entering the disk we expect to get from looking at the adiabatic invariant. The final change is the factor $\kappa \approx 0.526$ which we naively would have set to $1$. This term reflects the fact that only approximately half (from the Virial theorem) of the energy we supply to the system will show up in its kinetic energy. The other half will show up in the potential of the puffed up disk which does not lead to greater velocity dispersion.   

\bibliography{BOM_bibliography}
 
\end{document}